\documentclass[12pt]{article}
\usepackage{epsfig,amssymb,amsfonts,amsmath,graphicx,cite}
\usepackage{graphicx}
\usepackage{subcaption}
\usepackage{color}

\newcommand{\be}{\begin{equation}}
\newcommand{\ee}{\end{equation}}
\newcommand{\bea}{\begin{eqnarray}}
\newcommand{\eea}{\end{eqnarray}}

\begin{document}

\title{New supersymmetry-generated complex potentials with real spectra}

\author{Oscar Rosas-Ortiz${}^1$, Octavio Casta\~nos${}^2$ and Dieter Schuch${}^3$\\
{\footnotesize ${}^1$Physics Department, Cinvestav, AP 14-740, 07000
M\'exico DF, Mexico}\\
{\footnotesize ${}^2$Instituto de Ciencias Nucleares, UNAM, AP 70-543, 04510 M\'exico DF, Mexico}\\
{\footnotesize ${}^3$Institut f\"ur Theoretische Physik, Goethe-Universit\"at Frankfurt am Main,}\\ {\footnotesize Max-von-Laue-Str. 1, D-60438 Frankfurt am Main, Germany}}

\date{}
\maketitle

\begin{abstract}{\footnotesize 
A new form to construct complex superpotentials that produce real energy spectra in supersymmetric quantum mechanics is presented. This is based on the relation between the nonlinear Ermakov equation and a second order differential equation of the Schr\"odinger type. The superpotentials so constructed are characterized by the Ermakov parameters in such a way that they are always complex-valued and lead to non-Hermitian Hamiltonians with real spectra, whose eigenfunctions form a bi-orthogonal system. As applications we present new complex supersymmetric partners of the free particle that are ${\cal PT}$-symmetric and can be either periodic or regular (of the P\"oschl-Teller form). A new family of complex oscillators with real frequencies that have the energies of the harmonic oscillator plus an additional real eigenvalue is introduced.}
\end{abstract}

\section{Introduction}

The factorization method is an outstanding algebraic technique used to analyze the spectral properties of exactly solvable potentials in quantum mechanics \cite{Inf51,And84,Mie04}. Although the first record of the method can be traced back to the Fock article on {\em configuration space and second quantization} published in 1932 \cite{Foc32}, and this was explicitly used by Dirac in the 1935 edition of his book \cite{Dir35} ``as a little stratagem to solve the spectral problem for the one-dimensional quantum oscillator'' \cite{Mie04}, the factorization is usually attributed to Schr\"odinger who wrote on the matter eight years after Fock \cite{Sch40}. The main idea is to reduce the second order differential form of the one-dimensional Hamiltonian $H= -\tfrac12 \frac{d^2}{dx^2} +V(x)$ to the product of two first order differential operators $A$, $B$, up to an additive constant $\epsilon$ (proper units are assumed); that is, $H=AB+ \epsilon$. In the harmonic oscillator case, the factors $A$ and $B$ are respectively the creation $a^{\dagger}$ and annihilation $a$ operators of the boson algebra $[a, a^{\dagger}]=aa^{\dagger} - a^{\dagger} a = 1$, with $\epsilon =\tfrac12 \equiv E_0$ the ground state energy. Following Fock, $H= N+\tfrac12$ corresponds to the energy of a single mode of the electromagnetic field and $N=a^{\dagger}a$ accounts for the number of photons in that mode. The method is also in the kernel of supersymmetric quantum mechanics, introduced by Witten as a `toy model' of what happens in quantum field theories \cite{Wit81}. This last connection arose renewed interest in the issues of factorization and facilitated the strengthening of supersymmetric quantum mechanics as an algebraic discipline that is interesting in itself \cite{And84b,And85,Bag00,Coo01} (for recent reviews see\cite{Mie04,Kha04,And04,Kha05,Suk05,Fer05,Fer10}). Nevertheless, the entire development of the factorization method ``shows chronological gaps and inconsistencies; the ideas emerge, disappear and re-emerge again'' \cite{Mie04}.   

The present work addresses the problem of exploring new prospects to the factorization method. In addition to the supersymmetric treatments already established to construct non-Hermitian Hamiltonians with either real eigenvalues or a combination of real and complex eigenvalues (see e.g. \cite{Bay96,And99,Bag01,Can98,Fer03,Ros03,Ros07,Fer08a,Fer08b} and references quoted therein), we use here the solutions of the Ermakov equation to construct complex-valued potentials that share their spectrum with a given solvable real potential. Common in references \cite{Bay96,And99,Bag01,Can98,Fer03,Ros03,Ros07,Fer08a,Fer08b} is 
the idea of considering the eigenvalue equation $Hu = \epsilon u$, with $u(x)$ not necessarily normalizable in $\mbox{Dom}(V) \subseteq \mathbb R$, as a seed repository of the Darboux transformations $\widetilde V = V  -2\tfrac{d^2}{dx^2}(\ln u)$. The latter allow the construction of new solvable potentials $\widetilde V$ in terms of the solutions of a given exactly solvable potential $V$ and the `superpotential' $\beta = -\tfrac{d}{dx}(\ln u)$. Thus, $\widetilde V = V + 2\tfrac{d \beta}{dx}$ can be constructed such that $\mbox{Dom} (\widetilde V) = \mbox{Dom}(V)$ and either  $\sigma(\widetilde V) =\sigma(V)$ or $\sigma(\widetilde V) =\sigma(V) \cup \{\epsilon \}$, with $\sigma(A)$ the spectrum of $A$. The former case is referred to as `strictly isospectral' and the second one as `almost isospectral'. In general, if both $V$ and $\beta$ are real-valued then the new potential $\widetilde V$ is real-valued. However, a large variety of Darboux deformations $\widetilde V$ can be achieved if the superpotentials $\beta$ are complex-valued functions \cite{Bay96,And99,Bag01,Can98,Fer03,Ros03,Ros07,Fer08a,Fer08b}. Indeed, if $V$ and $\beta$ are respectively real and complex-valued functions then necessarily $\widetilde V$ is complex-valued. If both $V$ and $\beta$ are complex then $\widetilde V$ will be complex in general. 

To our knowledge, the current methods for constructing a complex-valued superpotential include at least one of the following ingredients:

\begin{itemize}

\item[i)] 
A normalizable complex function $u$ belonging to the eigenvalue $\epsilon \in \mathbb C$ of the non-Hermitian potential $V \in \mathbb C$. Usually $\mbox{Dom}(V)=\mathbb R^+$ and $u(0)=0$ \cite{Bay96}, but it can be extended to $\mbox{Dom}(V)=\mathbb R$ and the proper boundary conditions on $u$.

\item[ii)] 
A complex linear combination $\beta = f+ig$ of real functions $f$ and $g$ that satisfy a set of four Riccati-like equations connecting the real and imaginary parts of the non-Hermitian potentials $V$ and $\widetilde V$ \cite{And99,Bag01}. 

\item[iii)] 
A complex linear combination $u = u_1 + c u_2$, $c \in \mathbb C$, of the two fundamental eigen-solutions $u_1, u_2$, associated to the eigenvalue $\epsilon \in \mathbb R$ of the Hermitian potential $V \in \mathbb R$. Clearly $W(u_1,u_2) = u_1 \tfrac{du_2}{dx}-\tfrac{du_1}{dx} u_2 \neq 0$ \cite{Can98}.

\item[iv)] 
A complex analytic function $u$ belonging to the eigenvalue $\epsilon \in \mathbb C$ of the Hermitian potential $V \in \mathbb R$ \cite{Fer03,Ros03}.

\item[v)] 
A Gamow-Siegert function $u$ associated to the resonance $\epsilon = E - \tfrac{i}{2} \Gamma$ of the Hermitian potential $V \in \mathbb R$ \cite{Ros07,Fer08a,Fer08b}. 

\end{itemize}
However, the identification of exactly solvable complex potentials that can be generated by factorization (supersymmetric, Darboux or intertwining approaches) is, by far, not yet over. The purpose of the present work is to show that new classes of complex potentials with real spectrum are still achievable from the factorization method. With this aim we first review the connection between the solutions of the eigenvalue equation\footnote{From now on we omit a global factor $1/2$ in the Hamiltonian to adopt the simpler mathematical notation $H=-\frac{d^2}{dx^2} +V(x)$. The expressions for $H$ and its eigenvalues as they usually appear in physics are easily recovered from our results by retrieving the $1/2$ factor. For instance, in mathematical notation the potential and energy eigenvalues of the harmonic oscillator can be read as $V(x)= x^2$ and $E_n= 2n+1$. The `physical expressions' are respectively given by $\frac{V(x)}{2}= \tfrac{x^2}{2}$ and $\tfrac{E_n}{2}=n+\tfrac12$.} $Hu=\epsilon u$ and the solutions of the Ermakov equation
\be
\frac{d^2 \alpha}{dx^2} = M \alpha + \frac{\lambda_0}{\alpha^3},
\label{erma1}
\ee
where $\lambda_0$ is a constant and $M$ may be a function of $x$. This last nonlinear equation dates from a theorem proved by Ermakov in 1880 which states that {\em if an integral of the equation 
\be
\frac{d^2 u}{dx^2} =Mu
\label{erma1a}
\ee 
is known, one can find an integral of the equation} (\ref{erma1}) and vice versa, {\em if a particular solution of} (\ref{erma1}) {\em is known, we can find the complete solution of} (\ref{erma1a}) \cite{Erm80}, \S20. Independent works by Pinney \cite{Pin50}, and Lewis and Riesenfeld \cite{Lew69}, motivated the application of the Ermakov equation in physics {\cite{Eli76,FerG03,Eul07,Car08,Sch08,Moy09,Cas13,Sch13,Rod14,Cru15} (for recent reviews see \cite{Esp10,Lea08}). Remarkably, the above mentioned relationship has been used to solve the quantum spectral problem of some potentials in terms of the so called Milne quantization condition \cite{Kor81,Kor82,Lee84}. On the other hand, the relationship between (\ref{erma1}) and the Riccati equation fulfilled by the real-valued superpotential $\beta = -\frac{d}{dx} (\ln \varphi_0)$, with $\varphi_0$ the ground-state wavefunction of a given one-dimensional potential, has been indicated in \cite{Kau96} to connect the conventional supersymmetric quantum mechanics with the Ermakov approach (as discussed above, in this case all the Darboux deformations $\widetilde V= V-2\frac{d^2}{dx^2} (\ln \varphi_0)$ are real-valued functions). In such a context, it has been shown that the Ermakov equation (\ref{erma1}) obeys Darboux (supersymmetric) covariance \cite{Iof03}. Here we consider stationary systems and show that, even for real potentials $V(x)$ and real eigenvalues $\epsilon$, the solutions of the Ermakov equation (\ref{erma1}) are useful to construct complex-valued solutions of $Hu=\epsilon u$. Using these last as `transformation functions' the superpotential $\beta = -\tfrac{d}{dx} (\ln u)$ can always be chosen to be complex-valued, so that all the new potentials $\widetilde V = V + 2\frac{d\beta}{dx}$ are complex.

Our interest in complex-valued potentials includes the analogy between the spatially varying refractive index $m(\vec r)$ and the potential $V(\vec r)$ in the scattering of light and quantum particles respectively (in general, $\vec r$ is a vector in $\mathbb R^3$). Such analogy is usually drawn in terms of the identification $m^2(\vec r) = 1- V(\vec r)/[E]$, with $[E]$ representing energy units \cite{Kok06}. If the scattering medium is absorbing, the refractive index is complex $m=n-ik$, with nonzero imaginary part $k >0$. Here $n$ stands for the refractive index relative to the medium and $k$ the absorption coefficient of the medium. We stress that $k$ could be as small as to be omitted from calculations in a first approach, but this is never exactly zero for actual scattering media (see e.g. the experimental data reported in \cite{Kok06} for the values of $k$). Within this analogy, the non-Hermitian potentials $V(\vec r) =V_0(\vec r) + i V_1(\vec r)$ appearing in the quantum ``description of the scattering and the compound nucleus formation by nucleons impinging upon complex nuclei'' \cite{Fes54} are referred to as `optical potentials' \cite{Mug04} (see also `atypical models' in \cite{Mie04}). The simplest model considers $V_0(r)$ as a radial rectangular well of depth $U$ and width $R$, and $V_1(r)$ as a constant fraction of $V_0$, namely $V_1= \zeta V_0$ with $\zeta >0$. Then $V_0$ can be interpreted as the average potential in the nucleus and $V_1$ as the parameter of absorption associated with the formation of the compound nucleus \cite{Fes54} (notice that $V_1$ is the negative constant $-\zeta U$ in $r<R$). Using the factorization method with $\epsilon=E -\tfrac{i}{2} \Gamma$, $\Gamma \geq 0$, it has been shown that the well potential $V_0(r)$ of \cite{Fes54} is indeed the seed of a family of optical potentials $\widetilde V(r) =V_0(r) + i\widetilde V_1(r)$, the imaginary part of which is not a constant but a function of the position that changes sign across different regions of $[0, +\infty)$ as $r$ increases \cite{Fer08a}. Thus, $\widetilde V_1$ describes local gain and loss of the stationary current density $j=2 \mbox{Im}(\psi \tfrac{d}{dr} \psi^*)$ as an $s$-wave $\psi(r)$ `propagates' along $\mathbb R^+$ (the same is true for a one-dimensional wave $\psi(x)$ propagating along $\mathbb R$ \cite{Fer08b}). Hereafter $z^*$ means the complex-conjugate of $z \in \mathbb C$. The factorization is also used to transform optical potentials with a deep real part $V_0$ into complex potentials with a shallow real part $\widetilde V_0$ by ``removing bound states'' of the Schr\"odinger equation associated to $V=V_0+iV_1$. The result is a class of supersymmetric ``phase-equivalent complex potentials'' which has been applied to study the $\alpha+ {}^{16}O$ scattering  \cite{Bay96}. The reader can find a thorough discussion of the capabilities of the factorization method to intertwine two different complex potentials that share the same real spectrum in \cite{And99}. In the same context, the ${\cal PT}$-symmetric potential $V(x)=-(ix)^N/2$ \cite{Ben98} has been used to generate families of complex potentials which have not the property of parity and time reversal invariance but admit a real spectrum \cite{Can98}. 

The organization of the paper is as follows. In Section~\ref{sec2} we use the Ermakov theorem to characterize the solutions of the eigenvalue equation $Hu=\epsilon u$, with $H$ and $\epsilon$ real, in a form that $u$ can be chosen either real or complex-valued. Consistently, these $u$-functions give rise to superpotentials that depend on the parameters of the Ermakov equation. In Section~\ref{sec3} we show that new exactly solvable complex potentials with real spectra can be generated by using such Ermakov-parameterized superpotentials. It is also shown that the left eigenfunctions of the new Hamiltonian $\widetilde H$ are not equal to its right eigenfunctions because this last operator is not self-adjoint. Therefore, although $H$ and $\widetilde H$ are supersymmetric partners, the conventional intertwining relationships are not enough to fully describe the properties of the vector spaces involved. It is then necessary to extend the operator algebra in order to include the intertwining relationships that connect $H$ with $\widetilde H^{\dagger}$, the adjoint of $\widetilde H$. Using the new (generalized) intertwining relationships we show that the eigenvectors of both, $\widetilde H$ and $\widetilde H^{\dagger}$, form a bi-orthogonal system. In Section~\ref{sec4} some examples are presented. These include diverse families of new complex periodic ${\cal PT}$-symmetric potentials as well as new complex-valued  P\"oschl--Teller potentials that have a single bound state (Section~\ref{ssec4.1}). We also obtain new} complex-valued potentials with the set of energies of the harmonic oscillator plus an additional real eigenvalue (Section~\ref{ssec4.2}). In all cases we use systematically a parametric version of the general solution of the Ermakov equation (\ref{erma1}), the derivation of which can be found in Appendix~\ref{appa}. The paper ends with some concluding remarks.

\section{Complex-valued wave functions and corresponding superpotentials}
\label{sec2}

Consider the one-dimensional Hermitian Hamiltonian (in proper units):
\be
H= -\frac{d^2}{dx^2} +V(x), \qquad \mbox{Dom}(V) \subseteq \mathbb R.
\label{ham1}
\ee
We are interested in solving the eigenvalue equation 
\be
-u''(x) + [V(x) - \epsilon ] u(x)=0, \quad \epsilon \in \mathbb R,
\label{schro1}
\ee
where $f'(x)$ stands for the derivative of $f(x)$ with respect to $x$ and $u$ is a complex-valued function which is not necessarily normalizable in $\mbox{Dom}(V)$. 

\subsection{Complex-valued wave functions}
\label{ssec2.0}

Let us take
\be
u(x) = u_0 \exp\left[ - \int^x \beta(y) dy \right]
\label{u2}
\ee
with $\beta(x)$ a complex-valued function to be determined and $u_0$ an arbitrary complex constant. The introduction of (\ref{u2}) in (\ref{schro1}) leads to the Riccati equation 
\be
-\beta'(x) + \beta^2(x) +[\epsilon -V(x)]=0.
\label{riccati1}
\ee
It is well known that the relationship (\ref{u2}) is reversed by the logarithmic equation
\be
\beta(x) = -\frac{d}{dx} \ln u(x),
\label{u2inv}
\ee
so that (\ref{u2inv}) `linearizes' the Riccati equation (\ref{riccati1}) by transforming it into the Schr\"odinger equation (\ref{schro1}). As we look for $\beta: \mbox{Dom}(V) \rightarrow \mathbb C$, it is useful to write $\beta(x) = \beta_R(x) + i \beta_I(x)$, with $\beta_R$ and $\beta_I$ real-valued functions to be determined. Then $u$ is rewritten as $u = \Phi e^{i\Xi}$, where the {\em amplitude} $\Phi(x)$ and {\em phase} $\Xi(x)$ are defined by the real-valued functions 
\be
\Phi(x) = \Phi_0 \exp \left(-\int^x \beta_R(y) dy\right), \quad \Xi(x) = - \int^x \beta_I(y) dy + \Xi_0.
\label{uing}
\ee
Here $\Phi_0$ and $\Xi_0$ are integration constants. A straightforward calculation shows that the stationary current density associated with $u$ is given by $j=2 \frac{d\Xi}{dx} \vert u \vert^2 = -2\beta_I \vert u \vert^2 \equiv v\rho$, with $v= -2\beta_I$ the corresponding `flux velocity' and $\rho$ the probability density \cite{Fer08b}. As a parameter, $v$ is useful in the study of the resonances belonging to short-range potentials in one dimension as well as for distinguishing between bound and scattering states \cite{Fer08a,Fer08b}. 

The real and imaginary parts of equation (\ref{riccati1}) lead to the coupled system
\begin{eqnarray}
-\beta'_R + \beta_R^2 - \beta_I^2 + \epsilon-V=0,
\label{ric1}\\[1ex]
-\beta'_I + 2 \beta_I \beta_R =0.
\label{ric2}
\end{eqnarray}
From (\ref{ric2}) one immediately gets
\be
\frac{d}{dx} \ln \beta_I = 2 \beta_R.
\label{ric2b}
\ee
This last equation suggest the ansatz
\be
\beta_R(x)= -\frac{d}{dx} \ln \alpha(x),
\label{super}
\ee
with $\alpha(x)$ a function to be determined. This ansatz provides a definite form for the solution of (\ref{ric2b}) in terms of the $\alpha$-function,
\be
\beta_I(x) = \frac{\lambda}{\alpha^2(x)},
\label{betaim}
\ee
where $\lambda$ is a constant obtained after integration. To get real-valued functions $\beta_R$ and $\beta_I$, it is enough to consider either real or purely imaginary $\alpha$-functions in Eqs.~(\ref{super}) and (\ref{betaim}), besides a real $\lambda$-constant in Eq.~(\ref{betaim}). Hereafter we assume that $\alpha$ and $\lambda$ are both real.

The introduction of (\ref{super}) and (\ref{betaim}) into (\ref{ric1}) leads to the nonlinear second order differential equation 
\be
\alpha''(x) = [V(x) -\epsilon]\alpha(x) + \frac{\lambda^2}{\alpha^3(x)},
\label{erma2}
\ee
which is the Ermakov equation (\ref{erma1}) for $M(x) =V(x) -\epsilon$ and $\lambda_0 =\lambda^2$ (for details on the possible values of $\lambda$ see the items listed below). The Equation~(\ref{erma2}) was actually discussed in the context of a nonlinear formulation of time-independent quantum mechanics by Reinisch \cite{Rei94} and its solution was used to construct complex solutions of the time-independent Schr\"odinger equation. From now on the dependence of $\alpha$ on the parameter $\lambda$ (equivalently  $\lambda_0$) is implicit. If necessary, we shall make it explicit by writing $\alpha(x; \lambda)$ instead of $\alpha(x)$. Using (\ref{super}) and (\ref{betaim}) in (\ref{uing}) yields
\be
\Phi(x) =\Phi_0 \alpha(x), \quad \Xi(x) -\Xi_0 = -\lambda \int^x \alpha^{-2}(y) dy.
\label{uing2}
\ee
Without loss of generality we take $\Phi_0 e^{i\Xi_0}=u_0=1$ to write
\be
u_{\lambda}(x) =  \alpha(x) \exp\left[ -i\lambda \int^x \alpha^{-2}(y) dy \right].
\label{u3}
\ee
Accordingly, the density $\rho_{\lambda} = \vert u_{\lambda} \vert^2 = \alpha^2$ and the flux velocity $v_{\lambda}=-2\beta_I=-2\lambda/\alpha^2$ associated to the complex-valued function (\ref{u3}) are position-dependent. Therefore, equations (\ref{ric1}) and (\ref{ric2}) can be interpreted as the stationary version of the Madelung's hydrodynamical equations \cite{Mad27}. 

Considering the relationship $\lambda^2=\lambda_0$ indicated above, one can write $\lambda_{\pm} = \pm \sqrt \lambda_0$, so that our function (\ref{u3}) is decoupled into the pair of particular integrals of the equation (\ref{erma1a}) reported by Ermakov in his theorem \cite{Erm80}:
\be
u_{\lambda_\pm}(x) =\alpha(x) \exp \left[ \mp i \sqrt{\lambda_0} \int^x \alpha^{-2}(y) dy \right].
\label{u3b}
\ee
We have the following classification of the $u$-functions:

\begin{itemize}
\item[(I)] 
If $\lambda_0=0$ then the pair (\ref{u3b}) is reduced to the same function $u_{\lambda_\pm=0} = \alpha$, and the Ermakov equation (\ref{erma2}) coincides with the Schr\"odinger equation (\ref{schro1}).

\item[(II)] 
If $\lambda_0 < 0$ then one obtains the purely imaginary numbers $\lambda_{\pm} = \pm i \sqrt{\vert \lambda_0 \vert}$, which  are not included in our approach although they lead to real-valued functions $u_{{\lambda}_{\pm}}$ (otherwise, the function $\beta_I$ in (\ref{betaim}) is not real-valued).

\item[(III)]
If $\lambda_0 >0$ then the $u$-functions are complex-valued.

\end{itemize}

On the other hand, we like to stress that (\ref{u3}) can be expressed also in the form
\be
u_{\lambda} = \Phi_0 \alpha \cos \left( \lambda \int^x \alpha^{-2}(y) dy + \Xi_0 \right) -i \Phi_0 \alpha \sin \left( \lambda \int^x \alpha^{-2}(y) dy + \Xi_0 \right),
\label{iof1}
\ee
where we have retrieved the arbitrariness of the constants $\Phi_0$ and $\Xi_0$. A simple inspection of (\ref{iof1}) shows that, up to the factor $\lambda$ in the argument of the $\sin$ function, the function $\mbox{Im}(u_{\lambda})$ can be connected with the solution, in amplitude-phase form, of the Schr\"odinger equation discussed in e.g. \cite{Iof03}. In this context, to identify the possible bound state energies $E_n$ of the system, one would impose the Milne quantization condition \cite{Kor81,Kor82,Lee84,Kau96,Iof03}:
\be
N(E) = \frac{\lambda}{\pi} \int_{-\infty}^{+\infty} \alpha^{-2}(x,E)dx \circeq n+1, \qquad n=0,1,\ldots,
\label{iof2}
\ee
to get $\mbox{Im}(u_{\lambda})=0$ as $\vert x \vert \rightarrow + \infty$. However, as our function (\ref{u3}) is complex-valued, the latter condition affects also the real part $\mbox{Re}(u_{\lambda})$, so that
\[
u_{\lambda} \rightarrow (-1)^{n+1} \Phi_0 \alpha \quad \mbox{as} \quad \vert x \vert \rightarrow + \infty.
\]
Thus, if one would like to describe normalized solutions to the spectral problem of $H$ with the $u$-functions, then the condition (\ref{iof2}) is not enough to cancel them at $ x = \pm \infty$; it is necessary also that $\alpha(x) \rightarrow 0$ as $\vert x \vert \rightarrow + \infty$. More precisely, in our approach, the cancellation of $\alpha$ at the edges of $\mbox{Dom}(V)$ is required to get regular $u$-functions addressed to represent physical states. Notice that one can proceed also by cancelling $\mbox{Re}(u_{\lambda})$ first and then determining the conditions to cancel $\mbox{Im}(u_{\lambda})$.

\subsection{New real and complex superpotentials}
\label{ssec2.1}

Assuming that the real function $\alpha(x)$ satisfies (\ref{erma2}) and $\lambda$ is a real number, the solution of the Riccati equation (\ref{riccati1}) is given by
\be 
\beta_{\lambda}(x) = -\frac{\alpha'(x)}{\alpha(x)} +i \frac{\lambda}{\alpha^2(x)}.
\label{beta}
\ee
In general, this last function is meromorphic in an open set of the complex plane with poles defined by the zeros of $\alpha$ in $\mbox{Dom}(V)$. Note that besides the conventional logarithmic term $(\ln \alpha)'$, our superpotential $\beta_{\lambda}$ includes an additive nonlinear term regulated by the $\lambda$-parameter that distinguishes the Ermakov equation (\ref{erma1}) from the linear second order differential equation (\ref{erma1a}). Taking into account the items (I)--(III) of the previous section, we identify three different classes of superpotentials.

\subsubsection{Conventional superpotentials ($\lambda=0)$}
\label{ssec2.1.1}

If $\lambda =0$ the superpotential (\ref{beta}) is a real-valued function $\beta_{\lambda=0} = -(\ln \alpha)'$ producing the Darboux transformations already studied in the conventional supersymmetric approaches. Thus, if $\lambda=0$ the derivation of the transformation functions $u$ in terms of the Ermakov equation is unnecessary. 

\subsubsection{Real superpotentials ($\lambda_{\pm}= \pm i \sqrt{\vert \lambda_0 \vert} \in \mathbb C)$}
\label{ssec2.1.2}

If $\lambda_0 <0$ then $\lambda$ is purely imaginary and the superpotential (\ref{beta}) decouples into the pair of real-valued functions
\be
\beta_{\lambda_{\pm}} = - (\ln \alpha)'  \mp \frac{\sqrt{\vert \lambda_0 \vert}}{\alpha^2}.
\label{breal}
\ee
However, as indicated in the item (II) above, this case is not included in our approach.

\subsubsection{Complex-valued superpotentials ($\lambda_{\pm}= \pm \sqrt{ \lambda_0} \in \mathbb R$)}
\label{ssec2.1.3}

If  $\lambda_0>0$ then $\lambda$ is a real number and the superpotential (\ref{beta}) decouples into the pair of complex-valued functions
\be
\beta_{\lambda_{\pm}} = - (\ln \alpha)'  \pm i \frac{\sqrt{ \lambda_0 }}{\alpha^2}.
\label{bcomp}
\ee
Observe that $\beta_{\lambda_-} = \beta_{\lambda_+}^*$. To avoid confusion with notation used in the previous case, hereafter we use $\beta_{\lambda_{\pm}}  = \beta_{\pm}$ if $\lambda_{\pm}= \pm \sqrt{ \lambda_0} \in \mathbb R$. Remarkably, the superpotentials (\ref{bcomp}) are complex-valued functions even if $\epsilon$ and $V$ in (\ref{riccati1}) are real. 

Note that the values of $\lambda$ producing real (complex) superpotentials at the same time give rise to real (complex) $u$-functions.

\section{New complex potentials with real spectrum}
\label{sec3}

Once we have constructed the complex-valued superpotential that solves the Riccati equation (\ref{riccati1}) in terms of the Ermakov equation (\ref{erma1}), we are able to factorize the initial Hamiltonian (\ref{ham1}). Let us consider the first order differential operators
\be
A = -\frac{d}{dx} + \beta_{\lambda}(x) \quad \mbox{and} \quad B=\frac{d}{dx} + \beta_{\lambda}(x),
\label{ab}
\ee
with $\beta_{\lambda}$ defined in (\ref{beta}). As  $\beta_{\lambda}$ satisfies (\ref{riccati1}), the product of $A$ with $B$ gives
\be
H = AB + \epsilon.
\label{ham2}
\ee
Hereafter the real parameter $\lambda$ will be different from zero, then the function $\beta_{\lambda}$ will be any of the complex-valued functions introduced in Section~\ref{ssec2.1.3}. The description of the transformations associated to the real superpotentials derived in Section~\ref{ssec2.1.1} for $\lambda =0$ can be done in terms of the already reported supersymmetric approaches. Then, it is a matter of substitution to verify that $H^{\dagger} = B^{\dagger} A^{\dagger} + \epsilon =H$, as $\beta_{\lambda}^*$ is necessarily a solution of (\ref{riccati1}). Thus, although $A$ and $B$ are not mutually adjoint they factorize $H$ in the ordering established by (\ref{ham2}) \cite{Ros03}. Now we reverse the order of the factors to get (the dependence of $\widetilde H$ on $\lambda$ is implicit):
\be
\widetilde H =BA + \epsilon = -\frac{d^2}{dx^2} + \widetilde V_{\lambda}(x), 
\label{ham3}
\ee
where the new potential $\widetilde V_{\lambda}(x)$ is a complex-valued function of $x$ that is regulated by the $\lambda$-parameter of the Ermakov equation,
\be
\begin{array}{rl}
\widetilde V_{\lambda}(x) & = V(x) + 2 \beta'_{\lambda} (x)\\[1ex]
&= \displaystyle V(x) -2 \frac{d^2}{dx^2} \ln \alpha (x) -  i 4 \lambda \frac{\alpha'(x)}{\alpha^3(x)}.
\end{array}
\label{darboux}
\ee
Notice that, in contrast with $H$, the operator $\widetilde H$ is not self-adjoint since $\widetilde V$ is complex. Indeed $\widetilde H^{\dagger} = A^{\dagger} B^{\dagger} + \epsilon = -\frac{d^2}{dx^2}+ \widetilde V_{\lambda}^* \neq \widetilde H$. On the other hand, it is straightforward to verify that the new potential $\widetilde V_{\lambda}(x)$ is associated to the nonlinear (Riccati) equation
\be
\widetilde V_{\lambda}(x) = \beta_{\lambda}'(x) + \beta_{\lambda}^2(x) + \epsilon.
\label{riccati2}
\ee
The comparison of equations (\ref{riccati1}) and (\ref{riccati2}) makes clear that, in general, they are not invariant under the change $\beta_{\lambda} \rightarrow -\beta_{\lambda}$. 

The factorizations (\ref{ham2}) and (\ref{ham3}) lead automatically to the intertwining relationships
\be
\widetilde H B = BH, \qquad HA = A \widetilde H,
\label{intertwin}
\ee
which, as is well known, are useful in constructing the eigenfunctions of $\widetilde H$ by the action of $B$ on the eigenfunctions of $H$ and vice versa, one gets the eigenfunctions of $H$ by acting $A$ on the eigenfunctions of $\widetilde H$. This point will be dealt with further in the next subsections.

\subsection{Eigenvalues and eigenfunctions of the new potentials}
\label{ssec3.1}

If $\epsilon \neq E$ and $\psi_E (x)$ is a normalized solution of the eigenvalue problem $H\psi_E = E\psi_E$, then $\widetilde \psi_E \propto B \psi_E$ is a solution of $\widetilde H \widetilde \psi_E = E \widetilde \psi_E$ that can be normalized for properly chosen $u$-functions (a simple calculation shows that $A$ reverses the action of $B$). There is only an additional eigen-solution $\widetilde \psi_{\epsilon}$ of the spectral problem of $\widetilde H$ that is not considered in the intertwining transformations (\ref{intertwin}). Moreover, it is associated to the eigenvalue $\epsilon \neq E$. Such a function is of the form $\widetilde \psi_{\epsilon} = c_{\epsilon} u_{\lambda}^{-1}$, with $c_{\epsilon}$ an arbitrary constant, and satisfies the condition $A\widetilde \psi_{\epsilon} =0$. It must be clear that the analytical properties of $\widetilde \psi_{\epsilon}$ are determined by the $u$-function. For instance, using (\ref{u3}) and $\lambda \in \mathbb R$ we have
\be
\vert \widetilde \psi_{\epsilon} \vert^2 =\left\vert \frac{c_{\epsilon}}{u_{\lambda}} \right\vert^2 =
 \frac{\vert c_{\epsilon} \vert^2}{\alpha^2}.
\label{norm}
\ee
Therefore, the zeros of $u_{\lambda}$ are poles of $\widetilde \psi_{\epsilon}$ and vice versa. As $\lambda \in \mathbb R$, the right hand side of (\ref{norm}) indicates that it is sufficient to analyze the zeros and divergences (if any exist) of $\alpha$ in order to get information of the square-integrability of $\widetilde \psi_{\epsilon}$. An $\alpha$-function that is free of zeros in $\mbox{Dom}(V)$ and diverges at the edges of $\mbox{Dom}(V)$, for example, is the ideal candidate to construct $\widetilde \psi_{\epsilon}$ with finite norm. Considering these remarks, if $\widetilde \psi_{\epsilon}$ is normalizable in $\mbox{Dom}(\widetilde V)=\mbox{Dom}(V)$, then it must be added to the set $\{ \widetilde \psi_E \}$, so that $\widetilde H$ is almost isospectral to $H$ because $\sigma(\widetilde H) = \sigma(H) \cup \{ \epsilon \}$. Otherwise $\sigma(\widetilde H) = \sigma(H)$. 

\subsection{Orthogonality in the new vector spaces}
\label{ssec3.2}

At this stage we would like to emphasize that although the functions $\widetilde \psi_E$ can be normalized, they are not mutually orthogonal. In this respect, if necessary, one can introduce a positive-definite scalar product under which the set $\{ \widetilde \psi_E \}$ is orthonormal (see for instance \cite{Mie03,Kre03}, `atypical models' in \cite{Mie04}, and references quoted therein). Another option arises by noticing that the algebraic relationships (\ref{intertwin}) lead to the maps
\be
B: {\cal H} \rightarrow \widetilde{\cal H} \quad \mbox{and} \quad A: \widetilde{\cal H} \rightarrow {\cal H},
\label{maps1}
\ee
where the Hilbert space ${\cal H} = \mbox{span}\{ \psi_E \}$ is the state-space\footnote{In this section we assume that the spectrum $\sigma (H)$ of $H$ is purely discrete $\sigma(H) = \sigma_d(H)$, so that the inner product $(\psi_E, \psi_{E'}) = \delta_{E E'}$, with $\delta_{ij}$ the Kronecker's delta, holds for any pair of energy eigenstates $\psi_E$ and $\psi_{E'}$. If scattering states are also present ($\sigma_c (H) \neq \emptyset$), then the Dirac normalization $(\psi_E, \psi_{E'})= \delta(E- E')$ must be considered for $E$ and $E'$ in $\sigma_c(H)$. However, some caution is required in order to extend the discussion of this section to the states belonging to the continuous spectrum of $H$ \cite{Sok06,And07,Sok07}.} associated with $H$ and $\widetilde{\cal H}$ is the representation space of  $\widetilde H$ induced by $B$. As the operator $\widetilde H$ is not self-adjoint we have formally a second pair of intertwining relationships 
\be
HB^{\dagger} = B^{\dagger} \widetilde H^{\dagger}, \qquad \widetilde H^{\dagger} A^{\dagger} = A^{\dagger} H.
\label{maps2}
\ee
In consequence, there is a pair of additional rules  
\be
A^{\dagger}: {\cal H} \rightarrow \overline{\cal H} \quad \mbox{and} \quad B^{\dagger}: \overline {\cal H} \rightarrow {\cal H}.
\label{maps3}
\ee
Here $\overline{\cal H}$ is the representation space of $\widetilde H^{\dagger}$ induced by $A^{\dagger}$. Therefore, the action of $A$, $B$, $A^{\dagger}$ and $B^{\dagger}$, on `ket' vectors differs from their action on `bra' vectors:
\be
\left\{
\begin{array}{l}
B \vert \psi_E \rangle \propto \vert \widetilde \psi_E \rangle\\[1ex]
A \vert \widetilde \psi_E \rangle \propto \vert \psi_E \rangle
\end{array}
\right.; \qquad
\left\{
\begin{array}{l}
\langle \psi_E \vert B^{\dagger} \propto \langle \widetilde \psi_E \vert\\[1ex]
\langle \widetilde \psi_E \vert A^{\dagger} \propto \langle \psi_E \vert
\end{array}
\right.,
\label{maps4}
\ee

\be
\left\{
\begin{array}{l}
A^{\dagger} \vert \psi_E \rangle \propto \vert \overline{\psi}_E \rangle\\[1.5ex]
B^{\dagger} \vert \overline{\psi}_E \rangle \propto \vert  \psi_E \rangle
\end{array}
\right.; \qquad
\left\{
\begin{array}{l}
\langle \psi_E \vert A \propto \langle \overline{\psi}_E \vert\\[1.5ex]
\langle \overline{\psi}_E \vert B \propto \langle \psi_E \vert
\end{array}
\right..
\label{maps5}
\ee
These last right--left action equations, together with the intertwining relationships (\ref{intertwin}) and (\ref{maps2}), give rise to two different forms of calculating the matrix elements of the diagonal operator $H-\epsilon$. Namely, for $E\neq \epsilon$, the matrix element
\be
\langle \psi_E \vert (H-\epsilon) \vert \psi_{E'}\rangle = (E-\epsilon) \langle \psi_E \vert \psi_{E'} \rangle
\label{maps6}
\ee
can be expressed as either 
\be
\langle \psi_E \vert AB \vert \psi_{E'}\rangle  \propto \langle \overline{\psi}_E \vert \widetilde \psi_{E'} \rangle \quad \mbox{or} \quad \langle \psi_E \vert B^{\dagger}A^{\dagger} \vert \psi_{E'}\rangle  \propto \langle \widetilde \psi_E \vert \overline{\psi}_{E'} \rangle.
\label{inner1}
\ee
As $\langle \psi_E \vert \psi_{E'} \rangle = \delta_{EE'}$, we can use (\ref{maps6}) and (\ref{inner1}) to fix the constants of proportionality indicated in (\ref{maps4})--(\ref{maps5}) as follows
\be
\vert \widetilde \psi_E \rangle = \frac{B \vert \psi_E \rangle}{\sqrt{E-\epsilon}}, \qquad \vert \overline{\psi}_E \rangle = \frac{A^{\dagger} \vert \psi_E \rangle}{\sqrt{E-\epsilon}}.
\label{inner2}
\ee
In this form the vectors $\vert \widetilde \psi_E \rangle$ and $\vert \overline{\psi}_E \rangle$ are bi-orthonormal
\be
\langle \overline{\psi}_E \vert \widetilde \psi_{E'} \rangle_{\widetilde {\cal H}} =\delta_{EE'},
\quad \langle \widetilde \psi_E \vert \overline{\psi}_{E'} \rangle_{\overline{\cal H}} = \delta_{EE'}, 
\quad \langle \overline{\psi}_E \vert \widetilde \psi_{E'} \rangle_{\widetilde {\cal H}}  = {\langle \widetilde \psi_{E'} \vert \overline{\psi}_E \rangle}^{\!*}_{\overline{\cal H}} \, ,
\label{inner3}
\ee
where the subscripts denote the representation space in which the kets are assumed as vectors and the bras as functionals. For instance, the operators $BA$ and $A^{\dagger} B^{\dagger}$ are defined to act on $\widetilde {\cal H}$ and $\overline{\cal H}$ respectively, then (remember $E\neq \epsilon$):
\be
\langle \overline{\psi}_E \vert BA \vert \widetilde \psi_{E'} \rangle_{\widetilde {\cal H}}= \langle \widetilde \psi_E \vert A^{\dagger} B^{\dagger} \vert \overline{\psi}_{E'} \rangle_{\overline{\cal H}} = \frac{1}{E-\epsilon} \langle \psi_E \vert (H-\epsilon)^2 \vert \psi_{E'} \rangle = (E-\epsilon) \delta_{EE'}.
\label{operarep}
\ee
The introduction of a bi-orthogonal system like $\{ \vert \widetilde \psi_E \rangle, \vert \overline{\psi}_E \rangle\}$ defines a connection between the representation spaces $\widetilde{\cal H}$ and $\overline{\cal H}$, with the inner product $\langle \cdot \vert \cdot \rangle$ in ${\cal H}$ as the mediator. Such an structure closes the spectral correlations induced by the algebraic rules (\ref{intertwin}) and (\ref{maps2}), and provides a mechanism to investigate the completeness of the eigenvectors of $\widetilde H$ (see for instance \cite{Bro14}). Nevertheless, a few precautions are necessary; to name some of the related challenges: (i) the physical interpretation of the products (\ref{inner1}) is not immediate if they lead to complex numbers \cite{Dir45}; (ii) the completeness inferred, at first sight, from the products (\ref{inner3}) would be not properly justified unless a series of conditions are satisfied \cite{Tan06}, see also \cite{Moz06} and  (iii) in contrast with the feasibility of finite-dimensional bi-orthogonal systems \cite{Bro14}, the existence and completeness of bi-orthogonal systems in the infinite-dimensional case is not obvious, specially if the continuum spectrum is present \cite{Bro14,Sok06,And07,Sok07}. Considering this last point, in particular, the resolution of  identity (and thereby the quantum averages of observables) could be threatened if vectors $\vert \widetilde \psi \rangle$ with {\em zero--binorm} $\langle \overline{\psi} \vert \widetilde \psi \rangle =0$ are present \cite{Sok06,And07,Sok07}. However, such a ``self-orthogonality'' phenomenon is neither accidental nor rare in physics \cite{Nar03}, so they must be considered in the approaches on the matter. Quite recently, for instance, some of the difficulties involving zero--binorm bound states in a continuous spectrum (just as possible divergences in calculating quantum averages) have been faced with a proper resolution of identity \cite{Sok06}. In what follows we pay attention to a pair of profiles associated with the rules (\ref{inner2})--(\ref{operarep}) of our approach. They correspond to the case when $\vert \overline{\psi}_E \rangle$ is the complex-conjugate of its dual $\vert \widetilde \psi_E \rangle$, and to the presence of the {\em missing state} $\vert \psi_{\epsilon}\rangle$ in our model. Major precisions concerning the resolution of identity and the calculation of quantum averages of observables will be presented elsewhere.

Using the resolution of identity $\mathbb I = \int_{\mathbb R} dx \vert x \rangle \langle x \vert$, the $x$-representation of the bi-product in $\widetilde {\cal H}$ is introduced as
\be
(\overline{\psi}_E, \widetilde \psi_{E'} )_{\widetilde {\cal H}} := \left. \langle \overline{\psi}_E \vert \mathbb I \vert \widetilde \psi_{E'} \rangle_{\widetilde {\cal H}}  \right\vert_{\mathrm{Dom}(V)} = \int_{\mathrm{Dom}(V)} \overline{\psi}_E^{\, *}(x) \widetilde \psi_{E'}(x) dx.
\label{inner3b}
\ee
Consistently, $(\overline{\psi}_E, \widetilde \psi_{E'} )_{\widetilde {\cal H}}  = {(\widetilde \psi_{E'}, \overline{\psi}_E)}^{\!*}_{\overline{\cal H}}= \delta_{EE'}$. In definition (\ref{inner3b}) we assume that the Darboux transformation (\ref{darboux}) produces potentials $\widetilde V_{\lambda}(x)$ that are bounded from below, just as $V(x)$ is. This can be achieved, for example, if the real part of $\widetilde V_{\lambda}(x)$ is bounded from below and it dominates over the imaginary one such that the ratio $\mbox{Im} \widetilde V_{\lambda}/\mbox{Re} \widetilde V_{\lambda}$ remains finite and sufficiently small at both edges of $\mbox{Dom}(V)$. Complex potentials with these properties have been glimpsed as a possible result of the supersymmetric quantum design at the time that they were classified as {\em soft non--Hermitian} \cite{And07}. As we are going to show in the next sections, this is precisely the type of potentials that arises from our model because $\mbox{Im} \widetilde V_{\lambda}$ behaves as $\alpha^{-3}$ whereas $\mbox{Re} \widetilde V_{\lambda}$ includes an additive term which behaves as $\alpha^{-2}$, with $\alpha$ a solution of the Ermakov equation (\ref{erma2}). Thus, the properly chosen $\alpha$-function allows the calculation of the inner product $(\overline{\psi}_E, \widetilde \psi_{E'})_{\cal \widetilde H}$ by preserving the boundary conditions on $\mbox{Dom}(V) \subseteq \mathbb R$, as it is indicated in (\ref{inner3b}). In turn, from (\ref{inner2}) the $x$-representation of $\vert \widetilde \psi_E \rangle$ and $\vert \overline{\psi}_E \rangle$ is respectively given by
\be
\widetilde \psi_E(x) =\tfrac{1}{\sqrt{E-\epsilon}} \left[ \psi_E'(x) + \beta(x) \psi_E(x)\right], \quad 
\overline{\psi}_E(x) =\tfrac{1}{\sqrt{E-\epsilon}} \left[ \psi_E'(x) + \beta^*(x) \psi_E(x)\right].
\label{inner4}
\ee
Then, if $\psi_E(x)$ is a real--valued function (as it occurs for the majority of the one-dimensional real potentials), we have $\overline{\psi}_E^{\, *}(x) = \widetilde \psi_E(x)$. Thus, the wave-function $\widetilde \psi_E(x)$ and its dual $\overline{\psi}_E(x)$ are complex-conjugate. In such a case the inner product (\ref{inner3b}) acquires the form
\be
({\widetilde \psi}_{E}^*, \widetilde \psi_{E'} ) = (\overline{\psi}_E, \overline{\psi}_{E'}^{\, *})  =\int_{\mathrm{Dom}(V)} \widetilde \psi_E(x) \widetilde \psi_{E'}(x) dx  =\int_{\mathrm{Dom}(V)} \overline{\psi}_E^{\, *}(x) \overline{\psi}_{E'}^{\, *}(x) dx= \delta_{EE'}.
\label{inner3c}
\ee
On the other hand, as indicated in the previous section, the missing state $\vert \widetilde \psi_{\epsilon} \rangle$ is in the kernel of the intertwining operator $A$. Therefore, given $\vert \overline{\psi}_E \rangle \in \overline{\cal H}$ with $E\neq \epsilon$, using (\ref{inner2}) one has
\be
\langle \overline{\psi}_E \vert \widetilde \psi_{\epsilon} \rangle = \frac{1}{\sqrt{E-\epsilon}}  \langle \psi_E \vert A \vert \widetilde \psi_{\epsilon} \rangle =0.
\label{misd1}
\ee
Hence, the missing state $\vert \widetilde \psi_{\epsilon} \rangle$ is bi-orthogonal to all the vectors $\vert \overline{\psi}_{E\neq \epsilon} \rangle \in \overline{\cal H}$. Now, let us investigate if there is a non-trivial vector $\vert \overline{\varphi} \rangle \in \overline{\cal H}$ such that it is bi-orthogonal to all the vectors $\vert \widetilde \psi_{E\neq \epsilon}  \rangle \in \widetilde {\cal H}$. From (\ref{inner2}) we get
\be
\langle \overline{\varphi} \vert \widetilde \psi_E \rangle = \langle \overline{\varphi} \vert B \vert \psi_E \rangle=0 \quad \Leftrightarrow \quad B^{\dagger} \vert \overline{\varphi} \rangle =0.
\label{misd2}
\ee
Therefore $\widetilde H^{\dagger} \vert \overline{\varphi} \rangle = \epsilon \vert \overline{\varphi} \rangle$, so that $\vert \overline{\varphi} \rangle \equiv \vert \overline{\psi}_{\epsilon} \rangle$ is dual to $\vert \widetilde \psi_{\epsilon} \rangle$ and is also bi-orthogonal to all the vectors $\vert \widetilde \psi_{E\neq \epsilon}  \rangle \in \widetilde{\cal H}$. Moreover, in contrast with (\ref{operarep}), which was calculated for $E\neq \epsilon$, now we have
\be
\langle \overline{\psi}_{\epsilon} \vert BA \vert \widetilde \psi_{\epsilon} \rangle_{\widetilde {\cal H}}= \langle \widetilde \psi_{\epsilon} \vert A^{\dagger} B^{\dagger} \vert \overline{\psi}_{\epsilon} \rangle_{\overline{\cal H}} =0.
\label{misd3}
\ee
This last result certifies that $\vert \widetilde \psi_{\epsilon}\rangle$ and $\vert \overline{\psi_{\epsilon}}\rangle$ are respectively eigenvectors of $\widetilde H$ and $\widetilde H^{\dagger}$ with eigenvlue $\epsilon \in \mathbb R$. To investigate the normalization of the missing state notice that $\overline{\psi}_{\epsilon}^{\, *}(x)  =\widetilde \psi_{\epsilon}(x) = c_{\epsilon} /u_{\lambda}(x)$, therefore
\be
\langle \overline{\psi}_{\epsilon} \vert \widetilde \psi_{\epsilon} \rangle = \int_{\mathrm{Dom}(V)} {\widetilde{\psi}_{\epsilon}}^2(x) dx = \int_{\mathrm{Dom}(V)} {\overline{\psi}^{\, *}_{\epsilon}}^2(x) dx.
\label{misd4}
\ee
In general, the bi-product (\ref{misd4}) depends on the analytic expression of the $\alpha$-function that defines the supersymmetric transformation. In particular, if $\mathrm{Dom}(V) \subseteq \mathbb R$ is symmetric with respect to the origin, the $\alpha$ function can be constructed such that (\ref{misd4}) is a finite real number, as it is shown in the next sections. In this form, the bi-orthonormal system $\{ \{\vert \widetilde \psi_{\epsilon} \rangle, \vert \widetilde \psi_E \rangle \}, \{ \vert \overline{\psi}_{\epsilon} \rangle, \vert \overline{\psi}_E \rangle \}\}$ underlies a viable mathematical structure for our model.

\section{Spectral design of complex-valued potentials}
\label{sec4}

Let $V$ be an exactly solvable potential such that $\mbox{Dom}(V) \subseteq \mathbb R$. That is, $V$ can be one-dimensional ($\mbox{Dom}(V) = \mathbb R$) as the free-particle, the modified P\"oschl--Teller or the harmonic oscillator potentials.  This can be also radial ($\mbox{Dom}(V) = \mathbb R^+$) as the Coulomb potential, or it can be defined even in a finite domain as this occurs for either the infinite square well, the trigonometric P\"oschl--Teller or the Scarf potentials. The Darboux transformation ({\ref{darboux}) and the intertwining relationships (\ref{intertwin}) allow for the construction of new exactly-solvable complex potentials $\widetilde V$ with a spectrum that is either identical to or almost the same as the one of the potentials indicated above. As we have seen, the conditions for mapping the physical wave functions of $V$ into physical wave functions of $\widetilde V$ depend on the analytical properties of the $\alpha$-function. Next we show the applicability of the method using the free-particle and the harmonic oscillator potentials as point of departure. In all cases we use a parametric form of the solution to the Ermakov equation that is convenient to simplify calculations, details on the derivation of this expression can be found in Appendix~\ref{appa}.


\subsection{Complex supersymmetric partners of the free particle potential}
\label{ssec4.1}

For the free particle potential $V_{free}(x)=0$ one can take $z=e^{ikx}$ and $v= e^{-ikx}$ as the fundamental integrals of (\ref{erma1a}) with $k^2 =\epsilon \geq 0$, and Wronskian $W(z,v)=-2ik$. Using Eqs.~(\ref{erma6}) and (\ref{erma7}) of Appendix~\ref{appa} (with $\lambda_0 = \lambda^2$), the parametric family of solutions to the corresponding Ermakov equation is given by
\be
\alpha(x) = \left[ae^{-2ikx} + ce^{2ikx} + 2 \sqrt{ac +\frac{\lambda^2}{4k^2}} \right]^{1/2}.
\label{free1}
\ee

\subsubsection{New complex periodic ${\cal PT}$-symmetric potentials}

As $k \in \mathbb R$ and $\lambda \in \mathbb R$, we have two immediate (but not exclusive) forms to get a real function $\alpha$. Namely, we can take either $a=c=1/2$ or $a=-c=i/2$. In the former case one has
\be
\alpha (x) = \left[ \cos(2kx) + \gamma_{\lambda} \right]^{1/2}, \quad \gamma_{\lambda} =  \sqrt{1+ \frac{\lambda^2}{k^2}}.
\label{alfree}
\ee
The superpotential (\ref{beta}) and the new potential (\ref{darboux}) are in this case
\be
\beta_{\lambda}(x) = \frac{k \sin (2kx) + i \lambda}{\cos(2kx) + \gamma_{\lambda}}, \quad \widetilde V_{\lambda}(x) =  \frac{4k^2[1+ \gamma_{\lambda} \cos(2kx)] + i4\lambda k \sin(2kx)}{(\cos(2kx) + \gamma_{\lambda})^2}.
\label{betafree}
\ee
Remark that $\gamma_{\lambda}>1$ because $\lambda \neq 0$, then the above expressions are free of singularities. The potential $\widetilde V_{\lambda}(x)$ has been depicted in Figure~\ref{fig0} for different values of $\lambda$ and $k$. In turn, the square amplitude $\rho_{\epsilon}(x; \lambda) =\vert \widetilde \psi_{\epsilon} \vert^2$ of the missing state 
\be
\rho_{\epsilon}(x; \lambda) = \frac{\vert c_{\epsilon} \vert^2}{\cos(2kx) + \gamma_{\lambda}}
\label{denfree}
\ee
is a legitimate probability density whenever the integral $\int_{\mathbb R} \rho_{\epsilon}(x; \lambda) dx$ is finite. Clearly, given $k\in \mathbb R$, there is no value of $\lambda \in \mathbb R$ such that this last condition is fulfilled. That is, using only the physical energies $E=k^2>0$ of the free particle to define the $u$-function we obtain a family of complex periodic potentials $\widetilde V_{\lambda}$ that have no bound states. However, it is well known that the spectra of periodic potentials are composed of allowed and forbidden bands of energy, the former associated with {\em Bloch functions} as eigen-solutions of the related Schr\"odinger equation \cite{Ree78}. Then the spectral analysis of the periodic potentials (\ref{betafree}) can be done in terms of the {\em transfer} (or {\em monodromy}) matrix \cite{Ree78} and will be reported elsewhere. On the other hand, the real and imaginary parts of $\widetilde V_{\lambda}(x)$ are clearly even and odd functions so that this last potential is invariant under the transformations $i \rightarrow -i$, $x \rightarrow -x$. In other words, given $\lambda$, the complex periodic potential $\widetilde V_{\lambda}(x)$ is a ${\cal PT}$ supersymmetric partner of the free particle potential. Compared with other expressions of periodic ${\cal PT}$-symmetric potentials (not obtained in supersymmetric approaches), our result (\ref{betafree}) could be classified as a generalization of the Scarf~I-trigonometric-type potentials reported in e.g. \cite{Lev00}. Interestingly, the periodicity of potentials like the ones defined in (\ref{betafree}) can be connected with the periodicity of optical crystals that are balanced in their gain and loss profiles \cite{Mid14}.

\begin{figure}[htb]
\centering 
\begin{subfigure}[b]{.3\linewidth}
\centering
\includegraphics[width=.99\textwidth]{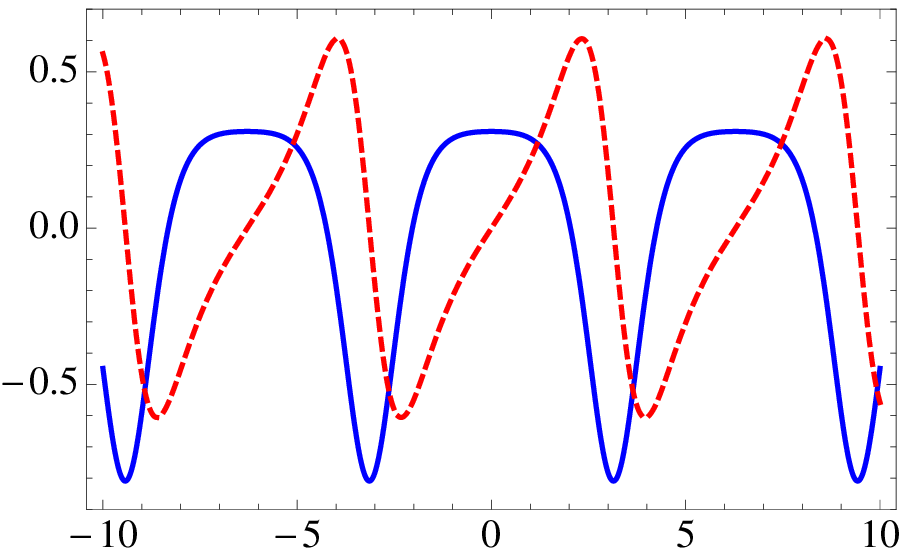}
\caption{$k=0.5$, $\lambda=1$}
\end{subfigure}
\begin{subfigure}[b]{.3\linewidth}
\centering
\includegraphics[width=.99\textwidth]{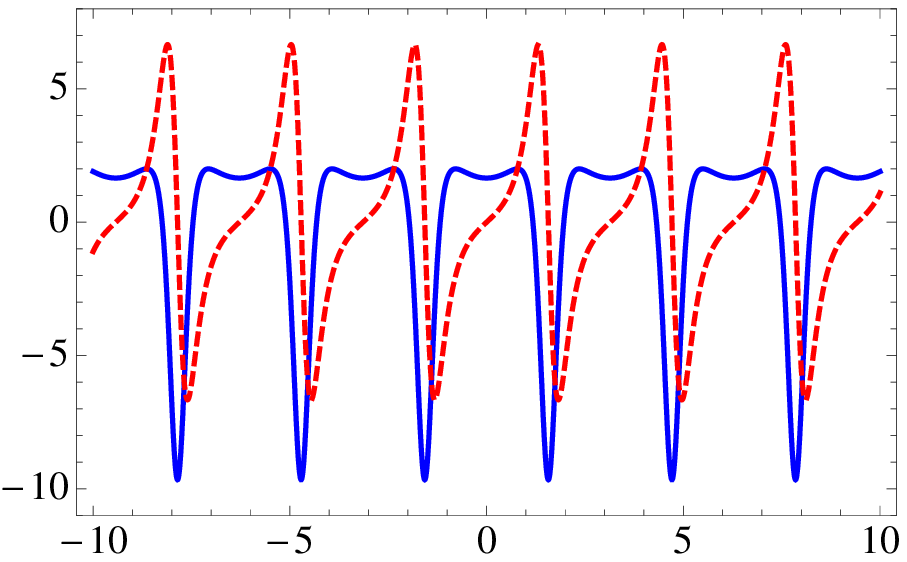}
\caption{$k=1$, $\lambda=1$}
\end{subfigure}
\begin{subfigure}[b]{.3\linewidth}
\centering
\includegraphics[width=.99\textwidth]{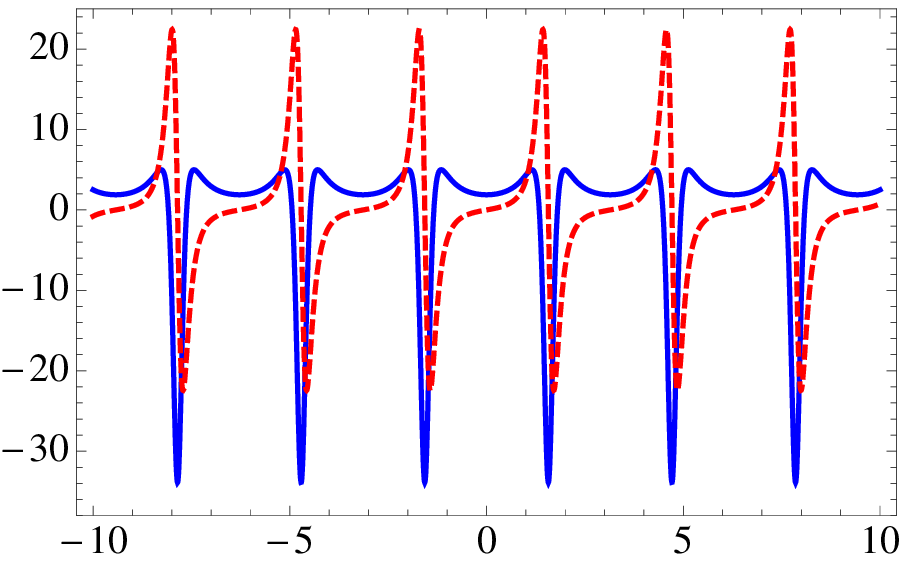}
\caption{$k=1$, $\lambda=0.5$}
\end{subfigure}
\caption{\footnotesize Members of the family of periodic complex potentials $\widetilde V_{\lambda}(x)$ introduced in (\ref{betafree}) for the indicated values of $\lambda$ and $k$. In all cases the continuos (blue) and dashed (red) curves correspond to the real and imaginary parts of the potential, respectively.}
\label{fig0}
\end{figure}

The other immediate option $a=-c=i/2$ produces
\be
\alpha (x) = \left[ \sin(2kx) + \gamma_{\lambda} \right]^{1/2},
\label{alfree2}
\ee
with results similar to the ones indicated above.

\subsubsection{New complex P\"oschl--Teller-like potentials}

To broaden the range of applicability of our approach let us consider unphysical eigenvalues ($E=k^2<0$) of the free particle potential. Using $k=i \tfrac{\kappa}{2}$, with $\kappa>0$, the periodicity of equations (\ref{alfree})--(\ref{alfree2}) is lost. Then, for the results associated with the superpotential (\ref{betafree}) we have the hyperbolic expressions\footnote{In this (hyperbolic) case, for simplicity in notation, we omit the {\em tilde} over the potential (\ref{hyper2}) and the wave function of the missing state (\ref{eigenfree}). Other functions as $\alpha$, $\beta_{\lambda}$ or $\rho_{\epsilon}$ do not include different notation.}:
\be
\alpha (x) = \left[ \cosh(\kappa x) + \theta_{\lambda} \right]^{1/2}, \quad \beta_{\lambda}(x) = \frac{-\frac{\kappa}{2} \sinh (\kappa x) + i \lambda}{\cosh (\kappa x) + \theta_{\lambda}},  \quad \theta_{\lambda} =\sqrt{1- 4 \left(\frac{\lambda}{\kappa}\right)^2}.
\label{hyper1}
\ee
Note that the $\alpha$-function is real-valued for $\lambda \in[-\frac{\kappa}{2}, \frac{\kappa}{2}]$. The new potential and square amplitude of the missing state are given by
\be
V_{\lambda} (x)= - \, \frac{\kappa^2 [1 + \theta_{\lambda} \cosh (\kappa x) ]+ i 2\lambda \kappa \sinh(\kappa x)}{(\cosh (\kappa x) + \theta_{\lambda})^2}, \quad
\rho_{\epsilon}(x; \lambda) = \frac{\vert c_{\epsilon} \vert^2}{\cosh(\kappa x) + \theta_{\lambda}}.
\label{hyper2}
\ee
\noindent
In this case the amplitude $\rho_{\epsilon}$ is finite along the straight line for all the values of $\kappa >0$ and $\lambda \in [-\frac{\kappa}{2},\frac{\kappa}{2}]$}. Indeed, the constant $c_{\epsilon}$ that normalizes such an amplitude is given by
\be
c_{\epsilon} =\sqrt{ \frac{\lambda}{2\mbox{arctan} \left[ \frac{\kappa}{2\lambda} \left( 1-\theta_{\lambda}
\right)
\right]}}, \quad \mbox{with} \quad \lim_{\lambda \rightarrow 0} c_{\epsilon} =\sqrt{\frac{\kappa}{2}}.
\label{cm}
\ee
Therefore, given $\lambda$ and $\kappa$, the complex-valued potential $V_{\lambda}(x)$ defined in (\ref{hyper2}) has the single energy $E=-\frac{\kappa^2}{4}$ with eigenfunction
\be
\psi_{\epsilon} (x; \lambda) = c_{\epsilon} \alpha^{-1}(x) \exp \left\{ i \arctan\left[ \frac{\kappa}{2\lambda} \left( 1-\theta_{\lambda}
\right) \tanh \left( \frac{\kappa x}{2}
\right)
\right]
\right\}.
\label{eigenfree}
\ee
The function $V_{\lambda}(x)$ in (\ref{hyper2}) really represents a $\lambda$-parameterized family of complex-valued potentials that share the same single-point discrete spectrum $\sigma(H_{\lambda}) = \{ E=-\frac{\kappa^2}{4} \}$. In Figure~\ref{fig1}, subfigures (a) and (b), we show two different members of the family $V_{\lambda}(x)$ that share a single discrete energy $E=-1/4$ (i.e.  $\kappa=1$). The corresponding wave functions $\psi_{\epsilon} (x; \lambda)$ are included in subfigures (c) and (d). Note the sensitivity of the imaginary parts of both the potentials and the wave functions to the variation of $\lambda$, in all cases their absolute value increases with $\vert \lambda \vert$. The real parts, on the other hand, have a less important sensitivity to the changes of $\lambda$. As we are considering $\lambda \in \mathbb R$, we know that $\beta_-$ is the complex conjugate of $\beta_+$ (see Section~\ref{ssec2.1.3}). The latter property produces $V_-(x)=V_+^*(x)$ and $\psi_{\epsilon}(x; \lambda_-) = \psi_{\epsilon}^*(x; \lambda_+)$, so that the imaginary parts of $V_-(x)$ and $\psi_{\epsilon}(x; \lambda_-)$ are the negative version of the ones depicted in Figure~\ref{fig1}. Indeed, the former correspond to the mirror image of the latter and vice versa.

\begin{figure}[htb]
\centering 
\begin{subfigure}[b]{.24\linewidth}
\centering
\includegraphics[width=.99\textwidth]{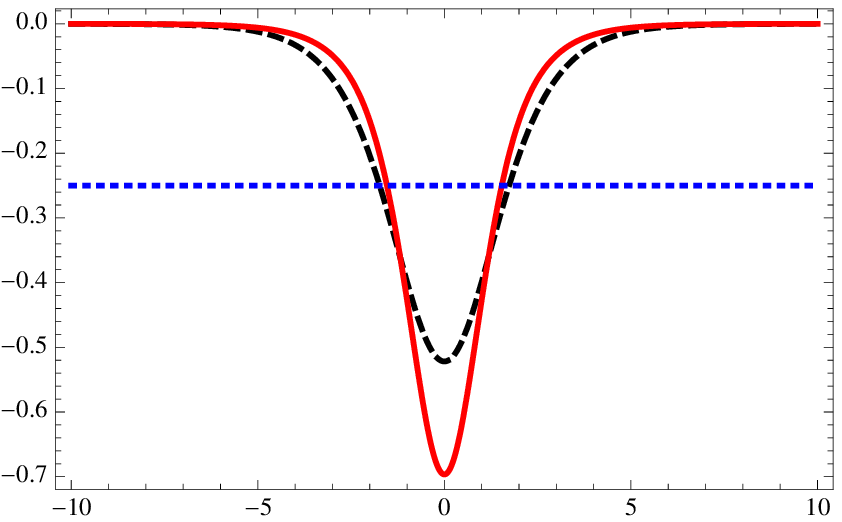}
\caption{$\mbox{Re}(V_{\lambda})$}
\end{subfigure}
\begin{subfigure}[b]{.24\linewidth}
\centering
\includegraphics[width=.99\textwidth]{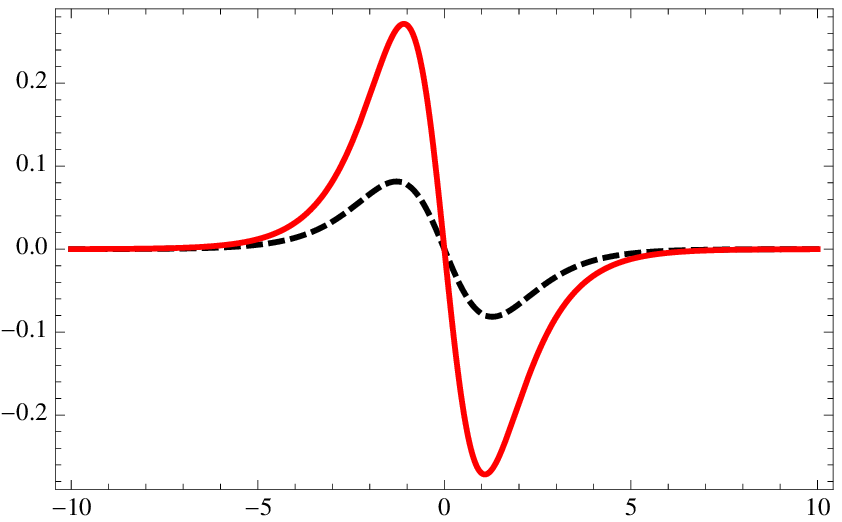}
\caption{$\mbox{Im}(V_{\lambda})$}
\end{subfigure}
\centering 
\begin{subfigure}[b]{.24\linewidth}
\centering
\includegraphics[width=.99\textwidth]{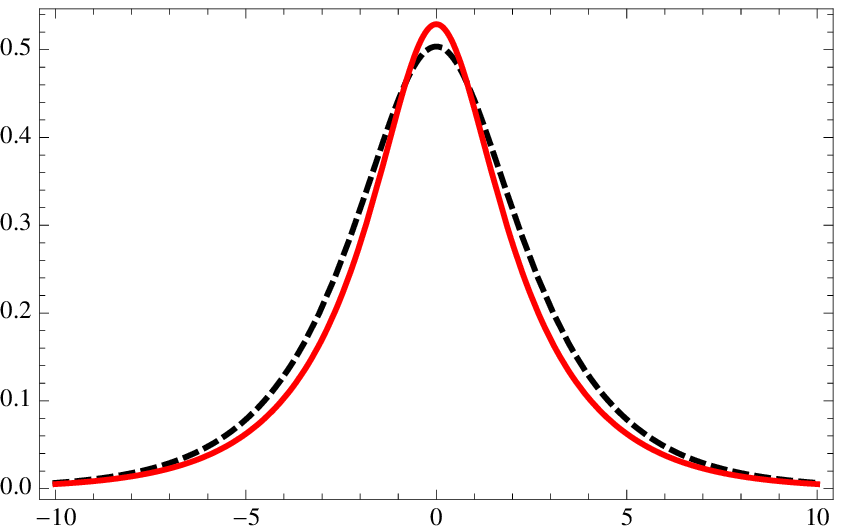}
\caption{$\mbox{Re}(\psi_{\epsilon})$}
\end{subfigure}
\begin{subfigure}[b]{.24\linewidth}
\centering
\includegraphics[width=.99\textwidth]{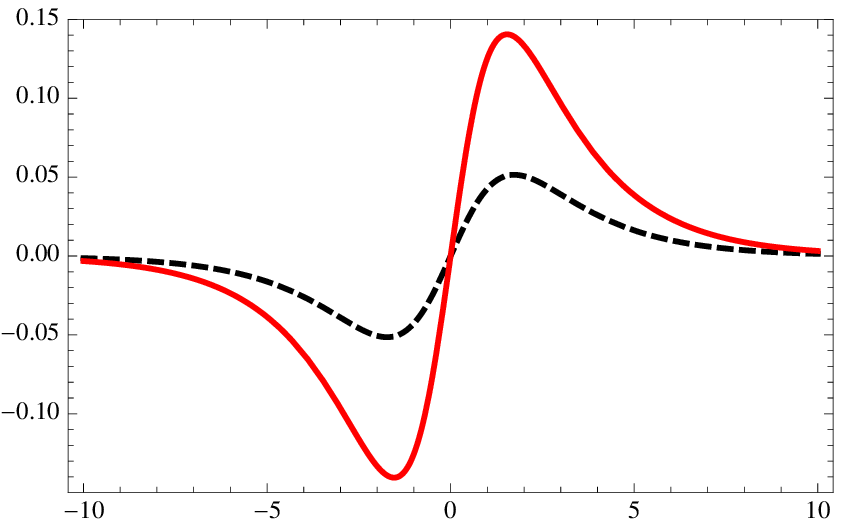}
\caption{$\mbox{Im}(\psi_{\epsilon})$}
\end{subfigure}
\caption{\footnotesize (a--b) Members of the family of complex-valued potentials $V_{\lambda}(x)$ defined in (\ref{hyper2}) for $\kappa =1$, $\lambda=0.2$ (black, dashed) and $\lambda =0.45$ (red, continuous). The dotted horizontal line, depicted in the real part of $V_{\lambda}$, refers to the single point discrete spectrum $\{ E=-1/4 \}$ that is shared by all the family members. (c--d) The corresponding wave functions $\psi_{\epsilon}(x;\lambda)$.}
\label{fig1}
\end{figure}

$\bullet$ ${\cal PT}${\bf -symmetry}

The symmetry indicated above is clear for $\psi_{\epsilon}$ since the argument of the exponential in (\ref{eigenfree}) is pure imaginary and the function $\arctan(x)$ is odd. Therefore, the change $\lambda \rightarrow -\lambda$ is equivalent to changing $i \rightarrow -i$. Concerning the potential $V_{\lambda}(x)$ introduced in (\ref{hyper2}), let us emphasize that the change $x \rightarrow -x$ reverts the $i$-transformation because $\alpha$ and $\cosh(x)$ are even functions while $\sinh(x)$ is odd. That is, the family of potentials reported in (\ref{hyper2}) is ${\cal PT}$-symmetric as the transformation $x \rightarrow -x$, $i \rightarrow -i$, leaves it invariant. The same holds for  $\psi_{\epsilon}$. Note however that the superpotential $\beta_{\lambda}$ becomes $-\beta_{\lambda}$ after such a transformation. We want to stress that our potentials (\ref{hyper2}) could be classified as a generalization of the Scarf~I-hyperbolic-type potentials reported in e.g. \cite{Lev00}. A similar classification holds for the potentials that are derived from the hyperbolic version of the superpotential (\ref{alfree2}).

$\bullet$ {\bf Supersymmetric P\"oschl--Teller profile}

The explicit analytical form of the new potentials (\ref{hyper2}) reveals a profile that is well known in the supersymmetric approaches associated to the free particle potential \cite{Coo01,Dia99,Mie00}. As a matter of fact, the modified P\"oschl--Teller potential $V_{PT}(x) = -\kappa^2 \cosh^{-2} (\kappa x)$ is one of the regular supersymmetric partners of $V_{free}(x)=0$ \cite{Mie00}. This last result gives rise to the identification of the delta well potential $V_{\delta}(x) = -\gamma_0 \delta(x)$, with $\gamma_0$ a properly chosen constant, as another supersymmetric partner of $V_{free}$ \cite{Dia99}. Here, the family of complex potentials (\ref{hyper2}) is also of the P\"oschl--Teller form. To confirm such statement remember that $\lambda=0$ must reduce our results to the conventional ones (Section~\ref{ssec2.1.1}). This last is clear since $\alpha(x; \lambda=0)= [\cosh(\kappa x) + 1]^{1/2} = \sqrt{2} \cosh (\frac{\kappa x}{2})$, which leads to the regular superpotential $\beta_{\lambda=0}(x) =-\frac{\kappa}{2} \tanh (\frac{\kappa x}{2})$ reported in \cite{Mie00}. Hence, the entire family $V_{\lambda}(x)$ is in the class of regular supersymmetric partners of the free particle potential and has a P\"oschl--Teller profile (compare with e.g. \cite{And99,Bag01}).

\begin{figure}[htb]
\centering 
\begin{subfigure}[b]{.24\linewidth}
\centering
\includegraphics[width=.99\textwidth]{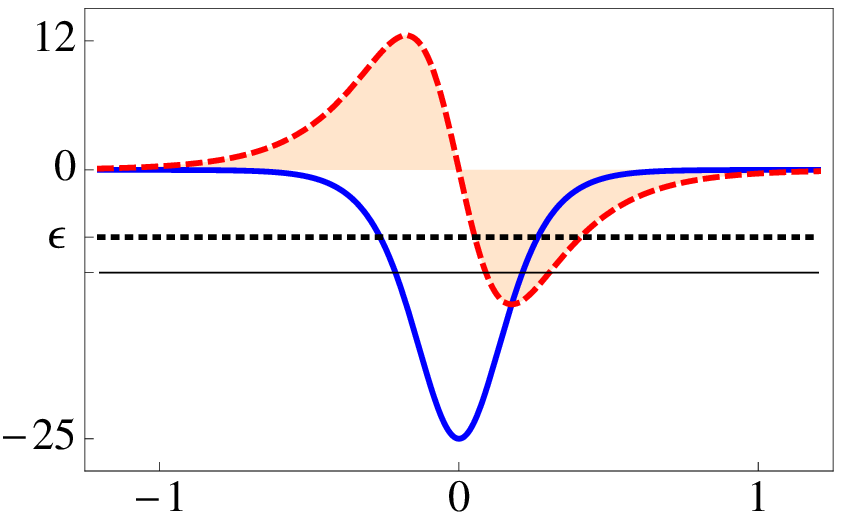}
\caption{$V_{\tfrac{\kappa}{2}}, \,\, \kappa =5$}
\end{subfigure}
\begin{subfigure}[b]{.24\linewidth}
\centering
\includegraphics[width=.99\textwidth]{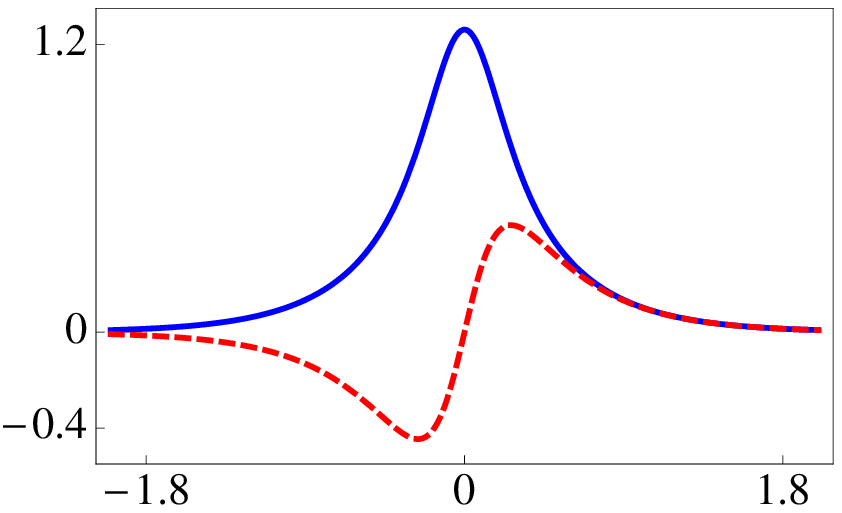}
\caption{$\psi_{\epsilon}, \,\, \kappa =5$}
\end{subfigure}
\centering 
\begin{subfigure}[b]{.24\linewidth}
\centering
\includegraphics[width=.99\textwidth]{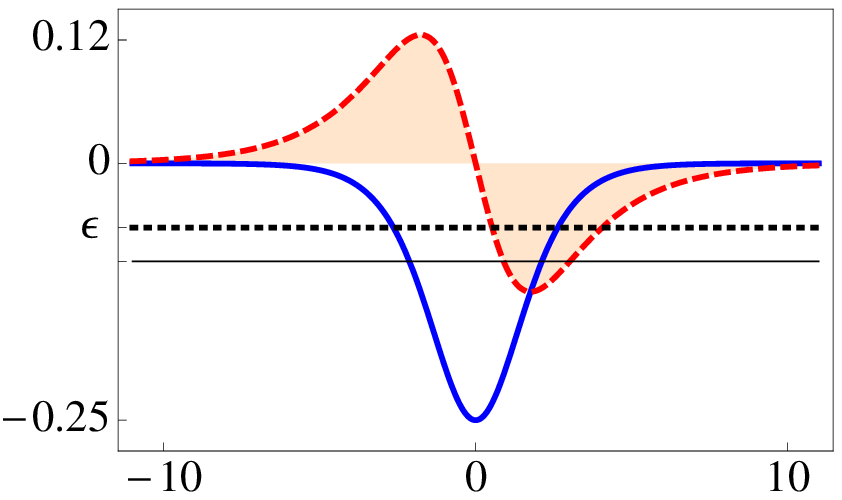}
\caption{$V_{\tfrac{\kappa}{2}}, \,\, \kappa =\frac{1}{2}$}
\end{subfigure}
\begin{subfigure}[b]{.24\linewidth}
\centering
\includegraphics[width=.99\textwidth]{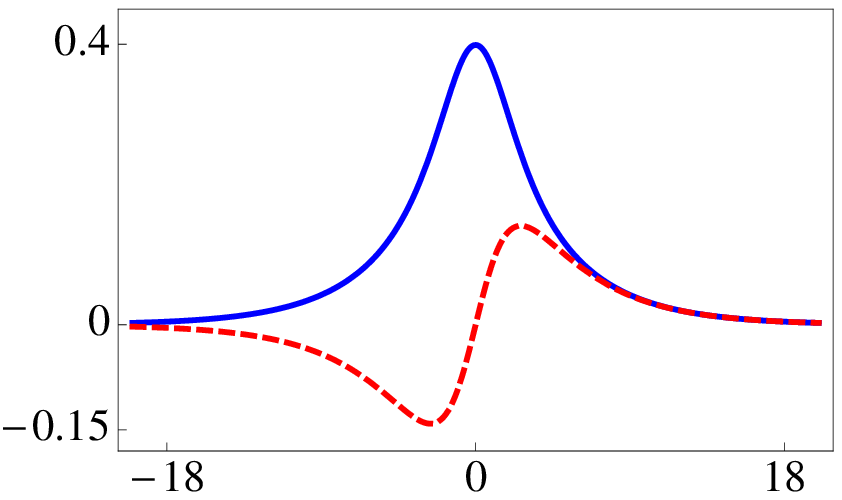}
\caption{$\psi_{\epsilon}, \,\, \kappa=\frac12$}
\end{subfigure}
\caption{\footnotesize Real (blue, continuous) and imaginary (red, dashed) parts of the complex-valued P\"oschl--Teller potential $V_{\tfrac{\kappa}{2}}(x)$ and the wave function $\psi_{\epsilon}$ for (a--b) $\kappa=5$ and (c--d) $\kappa=1/2$. The single discrete energy $\epsilon =-\tfrac{\kappa^2}{4}$ (dotted, black) of $V_{\tfrac{\kappa}{2}}(x)$ is less negative than the single energy $E_0 \approx 1.5 \epsilon$ (black) of the modified P\"oschl--Teller potential $V_{PT}(x)$. The imaginary part of the potential is shadowed to emphasize its gain and loss profiles.
}
\label{fig2}
\end{figure}

To give a particular example let us cancel the number $\theta_{\lambda}$ defined in (\ref{hyper1}). Then, using $\lambda = \frac{\kappa}{2}$ in (\ref{hyper2}), we arrive at the function
\be
V_{\tfrac{\kappa}{2}}(x) = -\frac{\kappa^2}{\cosh^2(\kappa x)} - i  \frac{\kappa^2 \sinh (\kappa x)}{\cosh^2(\kappa x)} = \left[ 1+i \sinh(\kappa x) \right] V_{PT}(x).
\label{free7}
\ee
This last expression includes the modified P\"oschl--Teller potential $V_{PT}(x)$ as its real part and has a single bound energy $\epsilon =-\tfrac{\kappa^2}{4}$, with normalized wave function
\be
\psi_{\epsilon} (x) = \left(\frac{\kappa}{\pi \cosh(\kappa x)} \right)^{1/2} \exp \!\! \left[ i \mbox{arctan} \left[\tanh \left( \frac{\kappa x}{2} \right)
\right]\right].
\label{free8}
\ee
The potential $V_{\tfrac{\kappa}{2}}(x)$ and the wave function $\psi_{\epsilon}(x)$ have been depicted for different values of $\kappa$ in Figure~\ref{fig2}.

$\bullet$ {\bf Optical model}

Using the optical model discussed in the introduction one can associate the imaginary part $ V_{I}(x) =\sinh (\kappa x) V_{PT}(x)$ of the potential $V_{\tfrac{\kappa}{2}}(x)$ with the absorption and emission of probability waves in $(0, +\infty)$ and $(-\infty,0)$ respectively (see Figure~\ref{fig2}). Notice that the rates of these processes coincide at $x$ and $-x$ respectively since the function $V_I$ is odd, so that the total probability is conserved. In Figure~\ref{fig2} we have shadowed (in light orange) the area from $V_I$ to the axis to emphasize the zones of $\mbox{Dom} (V_{\tfrac{\kappa}{2}})$ where this function works as a sink $(V_I <0)$ or as a source $(V_I >0)$ of probability waves. In this context, the complex potential $V_{\tfrac{\kappa}{2}}$ corresponds to a modified P\"oschl--Teller system $V_{PT}$ embedded in an environment $V_I$ which preserves the number of particles because the source and sink operate at the same rate in $(-\infty,0)$ and $(0, \infty)$ respectively. For scattering energies the function $V_{PT}(x)$ represents the average potential to which an incoming particle is subjected and $V_I(x)$ regulates the `capturing' and `releasing' of such a projectile by the scatterer in a vicinity of the maximum interaction. As $V_{\tfrac{\kappa}{2}}(x)$ is ${\cal PT}$-invariant, the analysis of the scattering properties of this potential could be completed by following e.g. Ref.~\cite{Can07}. 

On the other hand, it is well known that $V_{PT}(x)$ has a single bound state of energy $E_0 = -\tfrac{\kappa^2}{4} (\sqrt{5} -1)^2 \approx 1.5 \epsilon$ \cite{Dia99}. Then, as the single energy of $V_{\tfrac{\kappa}{2}}(x)$ is at $\epsilon= -\frac{\kappa^2}{4}$, the effect of the environment $V_I$ on the P\"oschl--Teller system $V_{PT}$ is the shifting of the single bound energy from $E_0 \approx 1.5 \epsilon$ to the less negative value $\epsilon=-\frac{\kappa^2}{4}$.

\subsection{Complex potentials with energies of the harmonic oscillator}
\label{ssec4.2}

As the fundamental solutions of the harmonic oscillator potential $V_{osc}(x)=x^2$ we use 
\be
z(x; \epsilon)={}_1F_1 \left[ \frac{1-\epsilon}{4}, \frac12, x^2 \right] e^{-x^2/2}, \quad v(x;\epsilon)=x {}_1F_1 \left[ \frac{3-\epsilon}{4}, \frac32, x^2 \right] e^{-x^2/2},
\label{nosc1}
\ee
with ${}_1F_1[a,c;z]$ standing for the confluent hypergeometric function \cite{Nie00,Ros03b,Olv10} and Wronskian $W(z,v)=-\tfrac12$. The introduction of these functions in the Equation~(\ref{erma6}) of Appendix~\ref{appa} gives rise to an $\alpha$-function that is parameterized by four numbers\footnote{Note that according to Eqs.~(\ref{abc})--(\ref{erma8}) of Appendix~\ref{appa}, the parameters $a$, $b$, and $c$ are respectively equivalent to the constants $c_0$ and $c_1$ that define the general solution of the Schr\"odinger equation (\ref{schro1}), and to the parameter $\lambda_0$ that characterizes the Ermakov equation (\ref{erma1}). These last parameters are used in Eqs.~(\ref{erma2a}) and (\ref{erma4}) to define the $\alpha$-function in Appendix~\ref{appa}. 
}: $a$, $b$, $c$ and $\epsilon$. This means we have at our disposal a vast source from which diverse families of complex potentials can be constructed with the energy spectrum of the harmonic oscillator. Indeed, as the $\beta_{\lambda}$-function (\ref{beta}) is always complex and its real and imaginary parts include the function $\alpha$ in the denominator, we have to choose the parameters $a$, $b$, and $c$ such that $\alpha$ is free of zeros in $\mbox{Dom}(\widetilde V_{\lambda}) = \mathbb R$ for a given $\epsilon$. This last prevents $\beta_{\lambda}$, $\widetilde V_{\lambda}$ and $\widetilde \psi_{\epsilon}$ of having singularities and opens the possibility of getting a well defined transformation of the initial wave functions into the new ones $\psi_E \rightarrow \widetilde \psi_E$. As indicated in Section~\ref{ssec3.1}, we will find that $\widetilde V_{\lambda}$ is strictly isospectral to $V_{osc}$ whenever $\widetilde \psi_{\epsilon}$ is not normalizable and $\beta_{\lambda}$ is free of singularities. However, our interest here is to manipulate the spectrum of the initial potential in order to design the spectrum of the new one. Henceforth, we look for a set of parameters $\{ a, b, c, \epsilon \}$ leading to regular complex superpotentials $\beta_{\lambda}$ and normalizable missing states $\widetilde \psi_{\epsilon}$. As an example let us consider the case $\epsilon =-1$, we have
\be
z(x) \equiv z(x,-1)=e^{x^2/2}, \quad v(x) \equiv v(x,-1)=\frac{\sqrt{\pi}}{2} e^{x^2/2} \, \mbox{Erf}(x)
\label{nosc2}
\ee
with $\mbox{Erf}(x)$ the error function \cite{Olv10}. Therefore, the $\alpha$-function (\ref{erma6}) can now be read as follows
\be
\alpha(x) =e^{x^2/2} \left[ a\, \mbox{Erf}^2 (x)  + b \, \mbox{Erf}(x) +  c  
\right]^{1/2},
\label{nosc3}
\ee
where unnecessary factors have been absorbed into $a$ and $b$. 

\subsubsection{Unbroken supersymmetry}
\label{ssec4.2.1}

Without loss of generality, let us assume that the parameters $a,b$ and $c$ in (\ref{nosc3}) are all non-negative. As $\mbox{Erf}(x) \in [-1,1]$ we realize that the $\alpha$-function is real and free of zeros if $c>\frac{b^2}{4a}$. Moreover, $\alpha \rightarrow e^{x^2/2}$ as $\vert x \vert \rightarrow +\infty$, so that $\frac{1}{\alpha^2}$ vanishes at the edges of $\mbox{Dom}(\widetilde V_{\lambda}) = \mathbb R$. Thus, for $\epsilon=-1$ we find that any set of non-negative parameters $\{a, b, c>\frac{b^2}{4a}\}$ produces regular superpotentials $\beta_{\lambda}$ and normalizable functions $\widetilde \psi_{\epsilon}$. In other words, the spectrum of the family of potentials
\be
\begin{array}{rl}
\widetilde V_{osc}(x; \lambda)  &= V_{osc}(x) + 2 \beta_{\lambda}'(x)\\[2ex]
& = \displaystyle x^2 -2 -2\frac{d}{dx} \left[ \frac{b+2a \mbox{Erf}(x) -i \sqrt{\pi}\lambda}{\sqrt{\pi} \, \alpha^2(x)}
\right]
\end{array}
\label{nosc4}
\ee
is the set of energies $\sigma(\widetilde V_{osc}) = \{ \epsilon =-1\} \cup \sigma(V_{osc}) \equiv \{ -1, 1,3, \ldots\}$. In the usual nomenclature one says that each member of the family $\widetilde V_{osc}(x; \lambda)$ is a {\em supersymmetric partner} of the harmonic oscillator potential $V_{osc}(x)$. Moreover, as the potentials belonging to such a family are almost isospectral to the harmonic oscillator, one also says that every pair $(\widetilde V_{osc}, V_{osc})$ corresponds to a system with {\em unbroken supersymmetry} \cite{Wit81,Bag00,Coo01}.

\begin{figure}[htb]
\centering 
\begin{subfigure}[b]{.3\linewidth}
\centering
\includegraphics[width=.99\textwidth]{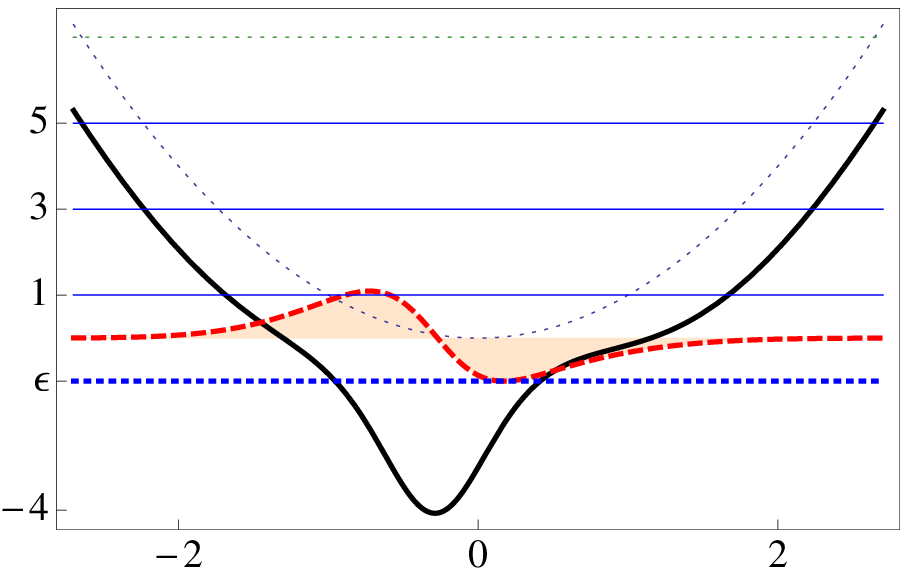}
\caption{$\widetilde V_{osc}(x; \lambda)$}
\end{subfigure}
\begin{subfigure}[b]{.3\linewidth}
\centering
\includegraphics[width=.99\textwidth]{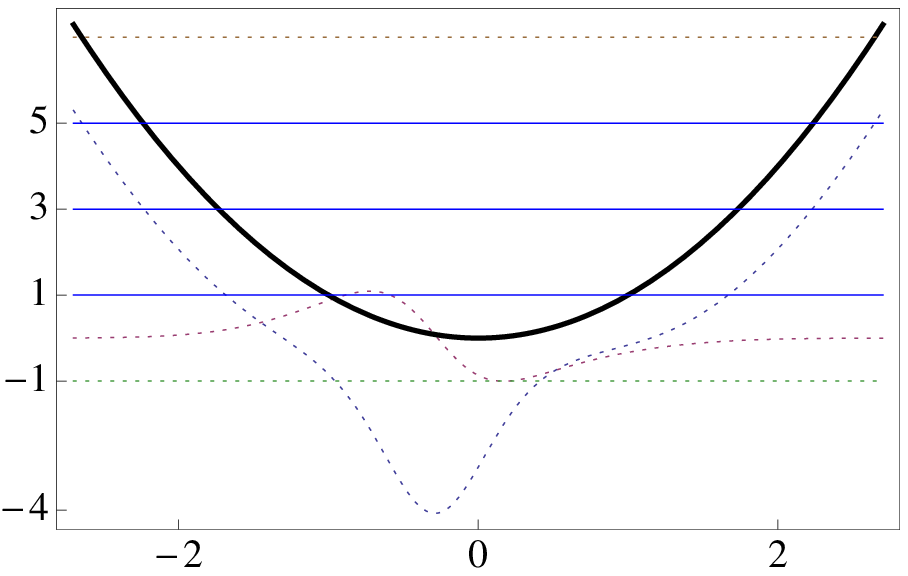}
\caption{$V_{osc}(x)$}
\end{subfigure}
\begin{subfigure}[b]{.3\linewidth}
\centering
\includegraphics[width=.99\textwidth]{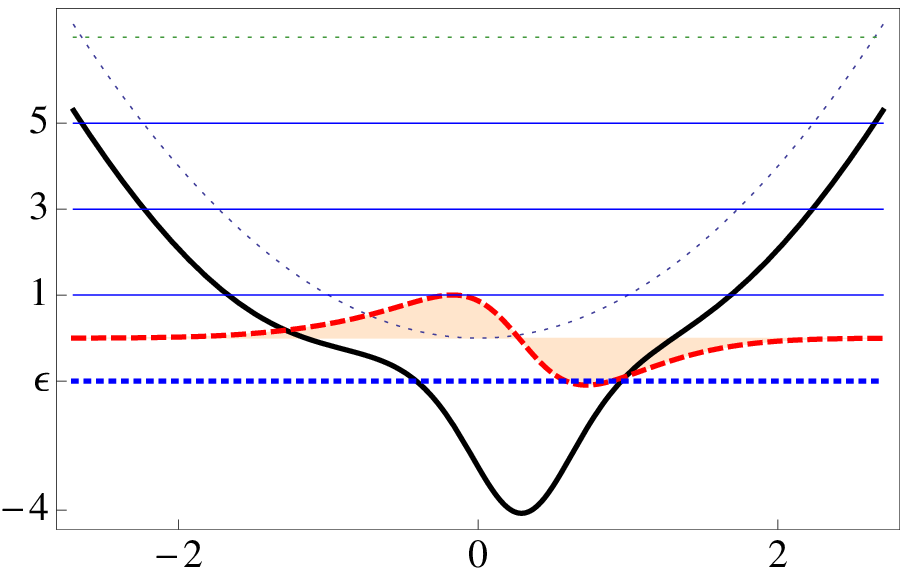}
\caption{$\widetilde V_{osc}^*(-x; \lambda)$}
\end{subfigure}
\caption{\footnotesize (a--b) The supersymmetric pair of oscillator potentials $(\widetilde V_{osc}, V_{osc})$ with unbroken supersymmetry generated with the parameters $a=\tfrac{\pi}{4}$, $b= \tfrac{\sqrt \pi}{2}$ and $c=1$ (equivalently, $\lambda = \frac{\sqrt{3\pi}}{8}$). The spectrum of the complex-valued potential $\widetilde V_{osc}$ is real and includes the energy $\epsilon =-1$ that is missing in the spectrum of the oscillator potential $V_{osc}$. The imaginary part of $\widetilde V_{osc}$ is in (dashed) red and shadowed to mark the zones in which it works as either a source or a sink. The ${\cal PT}$-transformation of $\widetilde V_{osc}$ produces the figure (c), so that this potential is not ${\cal PT}$-invariant. The potential $\widetilde V^*_{osc}(-x; \lambda)$ is, however, a member of a family of potentials for which $b \rightarrow -b$, so that it is also a supersymmetric partner of $V_{osc}$. }
\label{fig3}
\end{figure}

\begin{figure}[htb]
\centering 
\begin{subfigure}[b]{.24\linewidth}
\centering
\includegraphics[width=.99\textwidth]{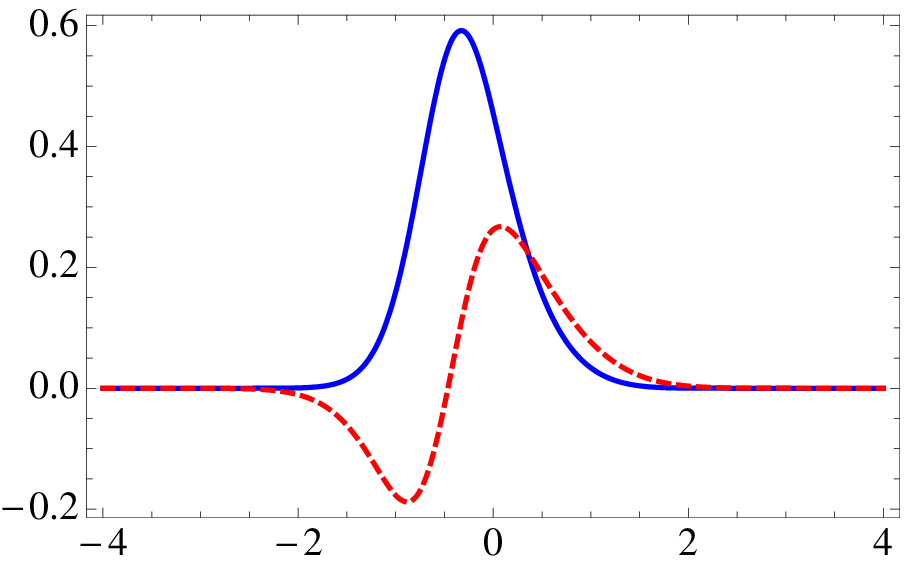}
\caption{$\epsilon \equiv \widetilde E_0= -1$}
\end{subfigure}
\begin{subfigure}[b]{.24\linewidth}
\centering
\includegraphics[width=.99\textwidth]{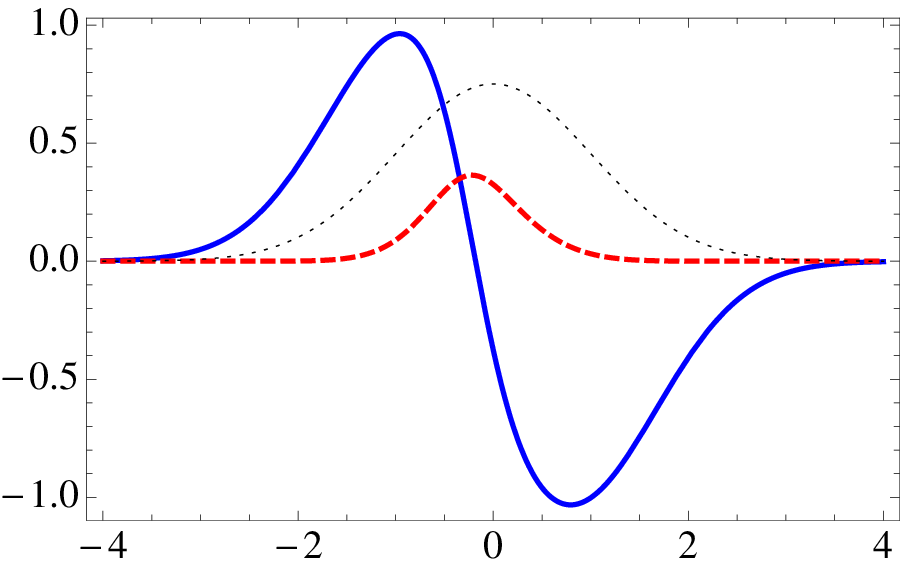}
\caption{$\widetilde E_1 (=E_0)=1$}
\end{subfigure}
\centering 
\begin{subfigure}[b]{.24\linewidth}
\centering
\includegraphics[width=.99\textwidth]{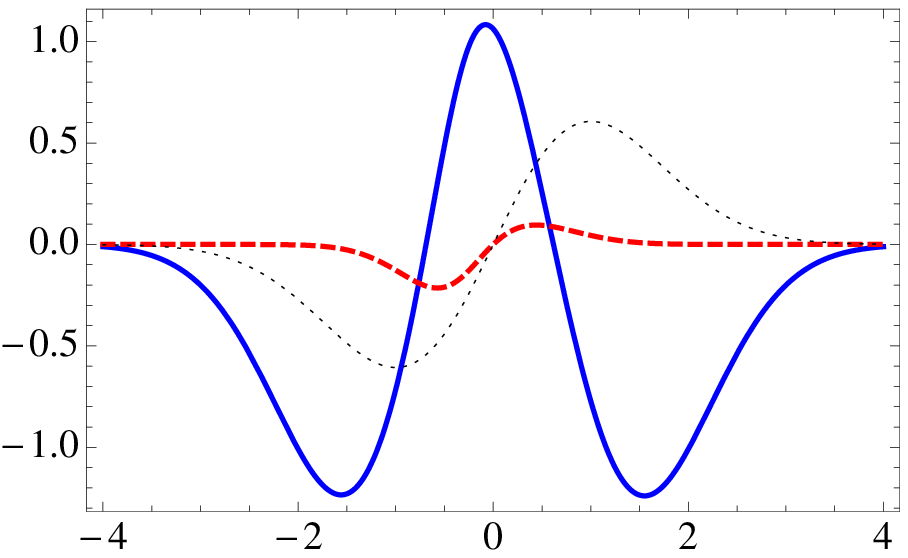}
\caption{$\widetilde E_2 (=E_1)=3$}
\end{subfigure}
\begin{subfigure}[b]{.24\linewidth}
\centering
\includegraphics[width=.99\textwidth]{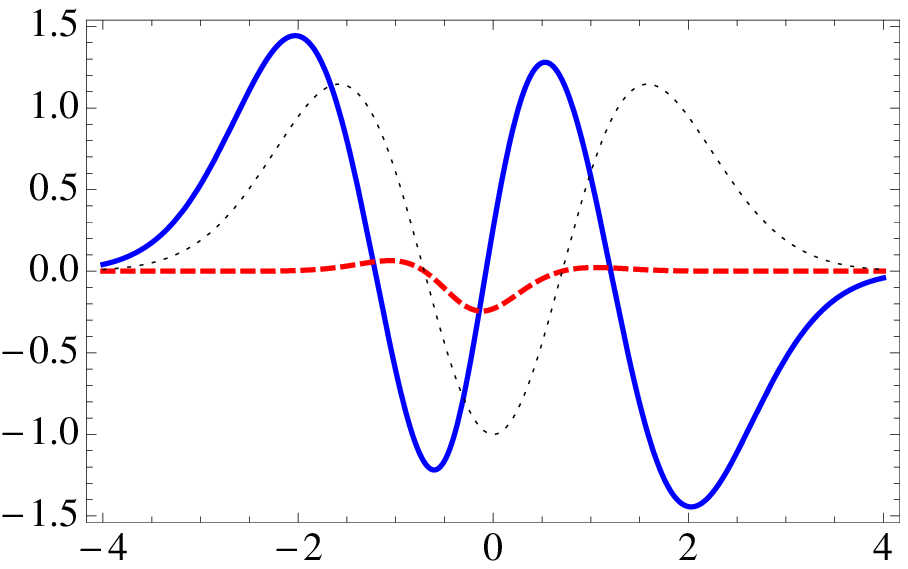}
\caption{$\widetilde E_3 (=E_2)=5$}
\end{subfigure}
\caption{\footnotesize The first four wave functions of the complex-valued potential $\widetilde V_{osc}(x; \lambda)$ shown in Figure~\ref{fig3} (a). In all cases the real part is in blue and the imaginary part in (dashed) red. Figures (b--d) include (in dotted, black) the corresponding oscillator wave functions. The wave function in (a) has no supersymmetric partner in the set of eigenfunctions of $V_{osc}(x)$ because the energy $\epsilon =-1$ is not included in the oscillator  spectrum. The wave functions associated to the ${\cal PT}$-transformed potential of Figure~\ref{fig3} (c) are immediately obtained from the ones shown here by using the same transformation.
}
\label{fig4}
\end{figure}

In Figure~\ref{fig3} (a--b) we show the behaviour of the supersymmetric pair $(\widetilde V_{osc}, V_{osc})$ for specific values of the parameters $a,b$ and $c$. These potentials share the same energy-levels $E=1,3,5,\ldots$, but the lowest one $\epsilon =-1$ is included only in the spectrum of $\widetilde V_{osc}$. Following Section~\ref{sec3}, one can depart from the spectrum of $V_{osc}$ to obtain the one of $\widetilde V_{osc}$ by the action of the operator $B$ (just as we have proceeded in the above section) and vice versa, the action of the operator $A$ on the wave functions of $\widetilde V_{osc}$ reverts the previous operation. The first four wave functions of the new potential $\widetilde V_{osc}$ are depicted in Figure~\ref{fig4} together with the oscillator ones. The missing state $\widetilde \psi_{\epsilon}(x)$, illustrated in Figure~\ref{fig4} (a),  has no supersymmetric partner in the set of eigenfunctions of $V_{osc}(x)$ because the spectrum of this last potential does not include the eigenvalue $\epsilon =-1$. Notice that in all the cases, the sign of the imaginary part of the potential and the wave functions changes around the minimum of the real part of the potential.

We want to remark that the `complex oscillator' that is usually reported in the literature is a variation of the conventional oscillator in which the frequency is allowed to be a complex number (see e.g. \cite{Jan86,Dav99,Fer15} and references quoted therein). As a consequence, the eigenvalues of the related non-Hermitian Hamiltonian are complex numbers, so that the spectrum is highly unstable under small perturbations\footnote{Despite this difficulty, a Lie-admissible theory has been applied in the study of the density matrices that represent the quantum states of such a complex oscillator \cite{Jan86} and numerical approximations have been developed to investigate the related resonances \cite{Dav99}. Moreover, the corresponding supersymmetric partners have been used to obtain solutions to the Painlev\'e IV equation \cite{Fer15}.}\cite{Dav99}. In contrast with the complex-frequency oscillator indicated above, our complex oscillators (\ref{nosc4}) are free of spectral difficulties because their frequencies and eigen-energies are purely real.

\subsubsection{Broken ${\cal PT}$-symmetry}
\label{ssec4.2.2}

For simplicity, in the previous section we have used a set of non-negative parameters $\{a, b, c>\frac{b^2}{4a}\}$ to get regular Darboux transformations of the oscillator spectral problem. Of course, this last set is not unique but serves to facilitate the identification of the new families of complex potentials with an exact solution to their spectral problem. Remarkably, such a set of parameters can be extended by the symmetries involved. For instance, the change $b \rightarrow -b$ preserves the analytical properties of the $\alpha$-function and gives rise to the potential depicted in Figure~\ref{fig3} (c). Thus, the change $b \rightarrow -b$ corresponds to the ${\cal PT}$-transformation $x\rightarrow -x$, $i \rightarrow -i$, so that it produces $\widetilde V_{osc}(x;\lambda) \rightarrow \widetilde V_{osc}^*(-x; \lambda)$. This last remark means that the new set of parameters $\{a, -b, c>\frac{b^2}{4a}\}$, with $a$ and $b$ non-negative numbers, represents another family of complex-valued potentials with a spectrum that is almost the same as the one of the harmonic oscillator. A member of this last family is precisely the potential in Figure~\ref{fig3} (c). Considering the results of the P\"oschl--Teller families discussed in Section~\ref{ssec4.1}, we realize that the finding of symmetries in the set of parameters obtained from the Ermakov equation to define the Darboux transformation is quite natural. However, in contrast with the periodic and P\"oschl--Teller cases discussed in the previous sections, here we have found a mechanism to produce complex potentials with real spectrum that are not invariant under ${\cal PT}$-transformations.

\section{Concluding remarks}

We have presented a new way of constructing complex superpotentials in supersymmetric quantum mechanics. The model is based on the relationship between the nonlinear equation (\ref{erma1}) introduced by Ermakov and the linear second order differential equation (\ref{erma1a}). The new aspect here is mainly represented by the transformation function $u_{\lambda}(x)$, used to generate the superpotential $\beta_{\lambda} = -(\ln u)'$,  because this is parameterized by the solutions of the Ermakov equation. In this form, the superpotential $\beta_{\lambda}(x)$  is for sure a complex-valued function, no matter if the initial potential $V(x)$ and the factorization constant $\epsilon$ are real in the Riccati equation (\ref{riccati1}) from which the supersymmetric transformation originates. As usual in the supersymmetric approaches, the method presented here includes both almost-- and strictly--isospectral partner potentials. In all cases our model ensures that the new complex potentials $\widetilde V_{\lambda}(x)$ are bounded from below in the context of the {\em soft non-Hermiticity} studied in \cite{And07}, and allows the construction of a bi-orthonormal system formed of the eigenvectors of $\widetilde V_{\lambda}(x)$ and the ones of its Hermitian conjugate. Moreover, in the almost--isospectral cases, the appropriate solution of the Ermakov equation gives rise to a vector state $\widetilde \psi_{\epsilon}(x)$ that is bi-orthogonal to the other eigenvectors $\widetilde \psi_E (x)$ of the complex potential $\widetilde V_{\lambda}(x)$. The function $\widetilde \psi_{\epsilon}(x)$ represents the ground state of  $\widetilde V_{\lambda}(x)$ and belongs to the real energy $\epsilon$, which is missing in the spectrum of the initial potential $V(x)$. Further properties of this pretty special `missing state' and the completeness of the  bi-orthonormal system associated with $\widetilde V_{\lambda}(x)$ are to be investigated.

We have illustrated the method by constructing new supersymmetric partners of the free particle potential that are either complex ${\cal PT}$-symmetric periodic potentials or complex ${\cal PT}$-symmetric regular potentials with a single discrete energy. The latter are of P\"oschl--Teller form, as expected for regular supersymmetric partners of $V_{free}(x)=0$. We have constructed also a family of complex-valued potentials $\widetilde V_{osc}(x)$ with the entire spectrum of the harmonic oscillator potential ($E=1,3,5,\ldots$), plus an additional energy at $\epsilon =-1$. All the members of this family are supersymmetric partners of the oscillator potential in such a way that each pair $(\widetilde V_{osc}, V_{osc})$ forms a supersymmetric system with unbroken supersymmetry. Remarkably, the new oscillator complex potentials $\widetilde V_{osc}$ have real frequencies and are not invariant under ${\cal PT}$-transformations. 

The procedure indicated above can be repeated at will, in order to produce new families of potentials. Moreover, it can be applied in the study of other potentials with domain of definition $\mbox{Dom}(V) \subset \mathbb R$, including the ones which have been already analyzed with the usual supersymmetric approaches.

\appendix
\section{Parametric solutions of the Ermakov equation}
\label{appa}
\numberwithin{equation}{section}
\setcounter{equation}{0}
\numberwithin{table}{section}
\setcounter{table}{0}

According to Ermakov \cite{Erm80}, the equation (\ref{erma1}) has the solution
\be
\alpha(x) = z(x)  \left[ \frac{\lambda_0}{c_1} + c_1 \left( \int^x z^{-2}(y) \, dy + \frac{c_0}{c_1} \right)^2
\right]^{1/2},
\label{erma2a}
\ee
where $c_1$ and $c_0$ are integration constants while $z(x)$ is a solution of the homogeneous equation (\ref{erma1a}). The structure of (\ref{erma2a}) allows to identify the function 
\be
v(x)= z(x) q(x), \quad \mbox{with} \quad q(x): = w_0 \int^x z^{-2}(y)\, dy
\label{erma4}
\ee
as the conventional expression for a second solution of (\ref{erma1a}) providing $z(x)$ as given \cite{Arf01}. It is assumed that the Wronskian $W(z,v) = zv'-z'v$ is a constant, $W(z,v) =w_0$, so that $z(x)$ and $v(x)$ are linearly independent. In consequence, $\alpha$ in (\ref{erma2a}) is reduced to the general solution of (\ref{erma1a}) if $\lambda_0 =0$, as Eqs. (\ref{erma1}) and (\ref{erma1a}) coincide for this value of $\lambda_0$. In terms of the fundamental pair $z$ and $v$, the solution (\ref{erma2a}) reads as follows
\be
\begin{array}{rl}
\alpha(x)  & =
z(x) \displaystyle\left[\frac{c_1}{w_0^2} q^2(x) + 2 \frac{c_0}{w_0} q(x) + \frac{\lambda_0 +c_0^2}{c_1}\right]^{1/2}\\[2ex]
& =
\displaystyle\left[ \frac{c_1}{w_0^2} v^2(x) +2 \frac{c_0}{w_0} v(x) z(x) + \left( \frac{\lambda_0 +c_0^2}{c_1} \right) z^2(x)
\right]^{1/2}.
\end{array}
\label{erma5}
\ee
The discriminant of the square root in the first line of this last equation is equal to $-4 \frac{\lambda_0}{w_0^2}$, therefore the solutions of (\ref{erma1}) are elements of the parameterized family
\be
\alpha(x) = \left[a v^2(x) + bv(x) z(x) + c z^2(x) \right]^{1/2}, 
\label{erma6}
\ee
where only two of the three parameters $a$, $b$ and $c$ are independent of each other. For
\be
a =\frac{c_1}{w_0^2}, \quad c= \frac{\lambda_0 +c_0^2}{c_1},
\label{abc}
\ee
it follows, with
\be
W(z,v)=w_0, \quad b^2-4ac=-4 \frac{\lambda_0}{w_0^2},
\label{erma7}
\ee
that
\be 
b=\pm 2 \frac{c_0}{w_0},
\label{erma8}
\ee
where $\lambda_0$ is still involved as a third parameter. The above expressions are consistent with the ones reported in e.g. \cite{Pin50,Eli76}  (see also \cite{Kor81,Kor82,Lee84,Kau96,Iof03}). Remark that any selection of the parameters $a,b$ and $c$ producing the linear combination $\alpha=z+\eta v$, with $\eta$ an arbitrary number, leads to $\lambda_0 =0$. According to our discussion in the previous sections we have the constraint $\lambda_0 >0$, so that $\alpha$ will be not the general solution of the Schr\"odinger equation (\ref{erma1a}) for the parameters $a,b$ and $c$ that we are considering in this work.

Another point of interest is to note that the expression 
\be
J= j_0 \left[ (u'\alpha - u \alpha')^2 + \lambda_0 \left( \frac{u}{\alpha} \right)^2 \right],
\label{J}
\ee
with the constant $j_0$ introduced for the sake of units, is $x$-independent. That is, $J' =0$. In this context, $J$ resembles the Ermakov dynamical invariant $I$. While $I$ is a constant of motion in the dynamical time-dependent case, the expression $J$ is here just a constant with respect to $x$. If $u$ is of the form (\ref{u3}), then such a constant is equal to zero. Therefore
\[
W(\alpha, u) = i \lambda_{\pm} \left( \frac{u}{\alpha} \right), 
\]
with $\lambda_{\pm} = \pm \sqrt{\lambda_0}$, see Section~\ref{ssec2.1}. This last result makes clear that $u$ and $\alpha$ are equivalent if $\lambda_0 =0$, as expected.

\section*{Acknowledgment}

This project commenced when D.S. visited the Instituto de Ciencias Nucleares, UNAM, and the Departamento de F\'{\i}sica, Cinvestav (M\'exico). D.S. expresses his gratitude to these institutions and particularly to Octavio Casta\~nos and Oscar Rosas-Ortiz for their kind hospitality. The support of CONACyT projects 238494 and 152574 is acknowledged. The authors acknowledge the Referee's comments addressed to include the results of Section~3.2.


\end{document}